\documentclass[fleqn,usenatbib]{mnras}

\usepackage[T1]{fontenc}

\DeclareRobustCommand{\VAN}[3]{#2}
\let\VANthebibliography\thebibliography
\def\thebibliography{\DeclareRobustCommand{\VAN}[3]{##3}\VANthebibliography}

\usepackage{graphicx}	
\usepackage{amsmath}	
\usepackage{amssymb}	
\usepackage{natbib}
\usepackage[varg]{txfonts}
\usepackage{graphicx}
\usepackage{wasysym}
\usepackage{color}
\usepackage{url}
\usepackage{tablefootnote}
\bibpunct{(}{)}{;}{a}{}{,}

\title[Mass-size Relations]{A possible signature of the influence of tidal perturbations in dwarf galaxy scaling relations}

\author[A. E. Watkins et al.]{
A. E. Watkins,$^{1}$\thanks{E-mail: a.emery.watkins@gmail.com (AEW)}
H. Salo,$^{2}$
S. Kaviraj$^{1}$
C. A. Collins$^{3}$
J. H. Knapen$^{4,5}$
A. Venhola$^{2}$
J. Rom\'{a}n$^{4,5,6}$
\\
$^{1}$Centre for Astrophysics Research, School of Physics, Astronomy and Mathematics, University of Hertfordshire, Hatfield AL10 9AB, UK\\
$^{2}$Space Physics and Astronomy Research Unit, University of Oulu, Pentti Kaiteran katu 1, FI-90014, Finland\\
$^{3}$Astrophysics Research Institute, Liverpool John Moores University, IC2 Building, Liverpool Science Park, 146 Brownlow Hill, Liverpool L3 5RF, United Kingdom\\
$^{4}$Instituto de Astrof\'{i}sica de Canarias, V\'{i}a L\'{a}ctea S/N, E-38205 La Laguna, Spain\\
$^{5}$Departamento de Astrof\'{i}sica, Universidad de La Laguna, E-38206 La Laguna, Spain\\
$^{6}$Kapteyn Astronomical Institute, University of Groningen, PO Box 800, 9700 AV Groningen, The Netherlands
}

\date{Accepted XXX. Received YYY; in original form ZZZ}

\pubyear{2022}

\begin{document}
\label{firstpage}
\pagerange{\pageref{firstpage}--\pageref{lastpage}}
\maketitle

\begin{abstract}
Dwarf galaxies are excellent cosmological probes, because their shallow potential wells make them very sensitive to the key processes that drive galaxy evolution, including baryonic feedback, tidal interactions, and ram pressure stripping.  However, some of the key parameters of dwarf galaxies, which help trace the effects of these processes, are still debated, including the relationship between their sizes and masses.  We re-examine the Fornax Cluster dwarf population from the point of view of isomass-radius--stellar mass relations (IRSMRs) using the Fornax Deep Survey Dwarf galaxy Catalogue, with the centrally located (among dwarfs) $3.63 \mathcal{M}_{\odot}$~pc$^{-2}$ isodensity radius defining our fiducial relation.  This relation is a powerful diagnostic tool for identifying dwarfs with unusual structure, as dwarf galaxies' remarkable monotonicity in light profile shapes, as a function of stellar mass, reduces the relation's scatter tremendously.  By examining how different dwarf properties (colour, tenth-nearest-neighbour distance, etc.) correlate with distance from our fiducial relation, we find a significant population of structural outliers with comparatively lower central mass surface density and larger half-light-radii, residing in locally denser regions in the cluster, albeit with similar red colours.  We propose that these faint, extended outliers likely formed through tidal disturbances, which make the dwarfs more diffuse, but with little mass loss. Comparing these outliers with ultra-diffuse galaxies (UDGs), we find that the term UDG lacks discriminatory power; UDGs in the Fornax Cluster lie both on and off of IRSMRs defined at small radii, while IRSMR outliers with masses below $\sim 10^{7.5} \mathcal{M}_{\odot}$ are excluded from the UDG classification due to their small effective radii.

\end{abstract}




\section{Introduction}

Dwarf galaxies are the most abundant kind of galaxy in the Universe \citep{schechter76, driver94, blanton05, baldry08, mcnaughtroberts14, kaviraj17}.  This, combined with their shallow potential wells, makes them excellent laboratories for galaxy evolution physics.  For example, the relative power of galactic superwinds from supernova, stellar and AGN feedback impacts structure much more efficiently in dwarf galaxies than in their massive counterparts \citep{lacey91, kay02, oppenheimer08, hopkins12, davis22}.  In gas-rich environments, ram pressure stripping \citep[RPS;][]{gunn72} might clean dwarfs of their gas as they infall, quenching their star formation (SF) and preventing further mass-buildup \citep[save, perhaps, in the deepest parts of their potential wells; e.g.,][]{lisker06, guerou15, venhola19, rude20}.  Tidal perturbations also leave noticeable marks on dwarfs, either through mass loss \citep[e.g., in the case of frequent high-mass-ratio encounters;][]{mcglynn90, moore98, kravtsov04, janz16, venhola19, jackson21}, increased vertical velocity dispersion \citep[e.g.,][]{moore98, mastropietro05}, or disruption \citep{mcglynn90, penarrubia08, koch12}.  Therefore, the history of a given environment's assembly is reflected by its dwarf population.

Galaxy clusters are the most massive gravitationally bound systems and thus host some of the most vibrant such assembly histories.  Indeed, their influence is clear even in populations of massive galaxies; many of the above-mentioned processes work to produce a clear morphology-density relation in clusters \citep{oemler74, dressler80, cappellari11}, in which the relative fraction of non-star-forming, mostly spheroidal early-type galaxies (ETGs) is significantly higher in the densest regions of clusters than in cluster outskirts or in the field.  Evidently, this translates as well to dwarf galaxies \citep[e.g.,][]{dressler80, ferguson90, grebel03, cote09, weisz11, habas20}, making early-type dwarfs \citep[dwarf ellipticals, dEs, and dwarf spheroidals, dSphs; e.g.,][]{kormendy09} the most abundant type of galaxy in cluster environments \citep[e.g.,][]{davies90, sabatini05, ferrarese12}.  Dwarf star-formation histories reflect environmental impact as well: for example, \citet{gallart15} found that dwarfs with rapid early SF followed by quenching exist in denser environments than dwarfs with continual, low-level SF.  Similarly, \citet{joshi21} showed that simulated dwarf satellites assemble most of their mass early, while dwarf centrals build theirs gradually. These environmental processes have a mass dependence, however: a high-mass dwarf will retain its gas longer than a low-mass dwarf in the same environment, simply because more massive systems are more robust to effects like ram pressure stripping (RPS) and tidal disruption \citep[e.g.,][]{simpson18, garrisonkimmel19}.

The term "dwarf" therefore refers not to a single monolithic population, but to a diverse population of gravitionally bound systems.  Such systems follow power-law correlations, also known as scaling relations, which connect many fundamental physical properties of galaxies---including dwarfs---giving insights into the galaxy formation and evolution processes.  For example, the correlation between the shapes of galaxies' light profiles---often parametrized by the S\'{e}rsic index \citep[$n$;][]{sersic63} or some measure of light concentration \citep[e.g.,][]{kent85, abraham94, bershady00}---with morphological type, bulge-to-disk ratio, surface brightness, and luminosity or stellar mass \citep[e.g.,][]{okamura84, caon93, bershady00, graham01, conselice03, ferrarese06, laurikainen10, holwerda14, paulinoafonso19} can help illustrate how global galaxy structure evolves.  Environment, intriguingly, seems to have comparatively little impact on light profile shape \citep[at least in the bright central regions of galaxies probed by large surveys or visible at high redshift; e.g.,][]{kauffmann04, vanderwel08, paulinoafonso19, martin19, jackson21}, suggesting that the structures traced by these parameters were in place early and have changed little over time \citep[e.g.,][]{szomoru12}.  Dwarfs follow similar such relations; however, in large enough populations, their concentration--stellar mass relation may deviate from that followed by more massive galaxies \citep[e.g.,][]{derijcke09, calderon15}, hinting at a kind of structural dichotomy emerging above a certain mass threshold \citep[but see, e.g.,][who instead argue that dwarfs and massive galaxies follow a curved continuum in structure, with no specific dichotomy]{graham03, ferrarese06, misgeld09, smithcastelli12}

Luminosity and stellar mass also correlate with galaxy size, whether that be the radius enclosing half the total light \citep[effective radius, $R_{\rm eff}$;][]{kauffmann03, shen03, trujillo04, franx08, vanderwel14, dosreis20}, or radii defined by specific values of mean azimuthal surface brightness \citep[e.g.,][]{devau91, haynes99, saintonge08, hall12, ouellette17, chamba20, trujillo20}.  Mass-size relations have been crucial to breaking degeneracies between different potential galaxy formation pathways.  Under the current concordance model of cosmology, $\Lambda$ cold dark matter ($\Lambda$CDM), galaxies build hierarchically through mergers \citep{white78, peebles80}, but the kinds of present-era galaxy structures resulting from such mergers depend on the types of mergers which dominate the hierarchical merging process \citep[e.g.,][]{hernquist91, naab09, szomoru12}.  Connecting galaxy masses, sizes, and dynamical properties like rotational velocity helps outline the distribution of angular momentum within dark matter haloes \citep[e.g.,][]{courteau07, dutton07, saintonge11, ouellette17}, a property tightly linked to accretion history.  Given the relationship discussed above between stellar mass and light profile shape, it is perhaps no surprise that light profile shape is, in turn, correlated with galaxy \citep[and, interestingly, globular cluster;][]{marchilasch19} size \citep[e.g.,][]{caon93, trujillo01, kelvin12, sanchezalmeida20}, ostensibly linking the bright central regions of galaxies (which form earliest) to their outermost extents (which form latest).

Scaling relations can thus help illuminate the processes that shape dwarf galaxy evolution across the varied and often violent environments within a galaxy cluster, particularly when contrasted with scaling relations followed by higher-mass galaxies.  In this paper, we compile and explore these various scaling relations for a sample of nearby galaxies spanning a wide range of stellar masses and morphologies, with a large sample of dwarfs taken from the Fornax Deep Survey Dwarf galaxy Catalogue \citep[FDSDC;][]{venhola18}, as well as a large sample of nearby dwarf and massive galaxies from the \emph{Spitzer} Survey of Stellar Structure in Galaxies \citep[S$^{4}$G;][]{sheth10}.  In Section \ref{sec:data}, we describe the observational datasets we use for this study.  In Section \ref{sec:methods}, we describe our methodology for deriving and analysing the various scaling relations we discuss.  In Section \ref{sec:results}, we present and discuss the scaling relations for this galaxy sample.  In Section \ref{sec:conclusions}, we discuss the implications of these scaling relations, with a focus on the outliers from them and their correlation with local environment measures.  Finally, we summarize our results in Section \ref{sec:summary}.

\section{Data}\label{sec:data}

For this study, we used both imaging data and derived data products from the Fornax Deep Survey \citep[FDS;][]{peletier20}, as well as the Spitzer Survey of Stellar Structure in Galaxies \citep[S$^{4}$G;][]{sheth10}, including both the original survey and the ETG extension \citep{sheth13, watkins22}.

The FDS is a joint survey using data from the VLT Survey Telescope (VST) Early-Type GAlaxy Survey (VEGAS, PIs: M. Capaccioli and E. Iodice) and the FOrnax Cluster Ultra-deep Survey \citep[FOCUS, PI: R. F. Peletier;][]{capaccioli15}.  Both surveys used the 2.6~m European Southern Observatory VST OmegaCAM \citep{kuijken02} to observe the Fornax Cluster and associated subgroup Fornax~A.  Observations covered a 26~deg$^{2}$ area centred on NGC~1399, in the $g^{\prime}$, $r^{\prime}$, and $i^{\prime}$ photometric bands and 21~deg$^{2}$ in the $u^{\prime}$ band, reaching substantial surface brightness depths: $5\sigma$ surface brightness limits $>27$ mag arcsec$^{-2}$ in all bands \citep{venhola18} or $3\sigma$ depths of $\gtrsim30$ mag arcsec$^{-2}$ on 10\arcsec$\times$10\arcsec \ scales, using the definition of limiting surface brightness proposed by \citet{trujillo16, roman20}.  Details regarding the observing strategy and data reduction for FDS are described by \citet{iodice16} and \citet{venhola18}.

For our analysis, we used the photometric data products produced by \citet{su21} for a sample of nearly 600 dwarf and massive galaxies, a revised version of the catalogue originally compiled by \citet{venhola18}.  On top of the 564 dwarfs from the FDSDC, this sample contains an additional 29 massive galaxies from the catalogues compiled by \citet{iodice19}, \citet{raj19}, and \citet{raj20}, and one additional dwarf galaxy not present in the FDSDC (FDS9\textunderscore0534).  Eight galaxies were also removed from the initial FDSDC due to duplication, for a final sample size of 582 galaxies.  Given their deeper limits and higher sensitivity for the mostly red galaxies in the Fornax Cluster, both versions of the catalogue only used the $g^{\prime}$, $r^{\prime}$, and $i^{\prime}$ images.  These data products include detailed photometric decompositions from which best-fit S\'{e}rsic profile parameters were derived, as well as quantities derived from radial surface brightness profiles and curves of growth, including effective radii and total integrated magnitudes.  Further, we derived isophotal and isomass radii from the surface brightness profiles produced by \citet{su21}, a process we describe in Sec.~\ref{sec:methods}.

To compare the Fornax dwarf population to the general, higher-mass galaxy population, we also use data from the S$^{4}$G, a magnitude-, size-, and distance-limited (total extinction-corrected blue magnitude $m_{B,\rm corr}<15.5$, isophotal diameter $D_{25}>1$\arcmin, and radial velocity $v<3000$km s$^{-1}$) survey of nearly 3000 nearby galaxies taken with the \emph{Spitzer} Space Telescope's \citep{werner04} Infrared Array Camera \citep[IRAC;][]{fazio04} in the 3.6$\mu$m and 4.5$\mu$m bands.  The combination of these bands make for nearly dust-free and nearly direct tracers of stellar mass \citep[e.g.,][]{pahre04, meidt14, driver16}.  Observations for the S$^{4}$G took place during the post-cryogenic phase of the \emph{Spitzer} mission, reaching surface brightness limits, on average, of $\mu_{3.6, \rm AB} \approx 25.5$ mag arcsec$^{-2}$ \citep[$3\sigma$ in 10\arcsec$\times$10\arcsec \ boxes;][]{salo15}, at a pixel scale of 0\farcs75~px$^{-1}$.  Details on the observing strategy and data reduction pipelines for the S$^{4}$G and its ETG extension survey are described by \citet{sheth10}, \citet{sheth13}, \citet{munozmateos15}, and \citet{watkins22}.

We used the photometric data products from Pipelines 3 \citep{munozmateos15} and 4 \citep{salo15} for both the original survey (2352 galaxies, mostly late-type) and the ETG extension \citep[465 newly observed ETGs;][]{sheth13, watkins22}.  Pipeline 3 produces radial surface brightness profiles and curves of growth, from which are derived effective radii, isophotal (easily converted to isomass) radii, total stellar masses, and concentration parameters.  Pipeline 4 produces detailed multi-component photometric decompositions using GALFIT \citep{peng02, peng10}, as well as best-fit single-component S\'{e}rsic profile parameters.  As the ETG extension sample is new, only single-component decompositions are currently available for these galaxies.  Given the survey's shallower depth compared with FDS, we re-derived isophotal radii from the Pipeline 3 radial profiles for this paper to smooth over the noise in the galaxies' low surface brightness outskirts, a process we describe in the following section.

With these two surveys combined, our sample contains surface photometry and GALFIT decompositions of 3400 galaxies spanning nearly six orders of magnitude in stellar mass across a wide range of environments in the nearby Universe.  The methods used to produce these photometric quantities are largely consistent across both surveys as well, making this combination of datasets uniquely suited to the kind of structural analysis we perform here.

\section{Methods}\label{sec:methods}

\begin{figure*}
    \centering
    \includegraphics[scale=1.0]{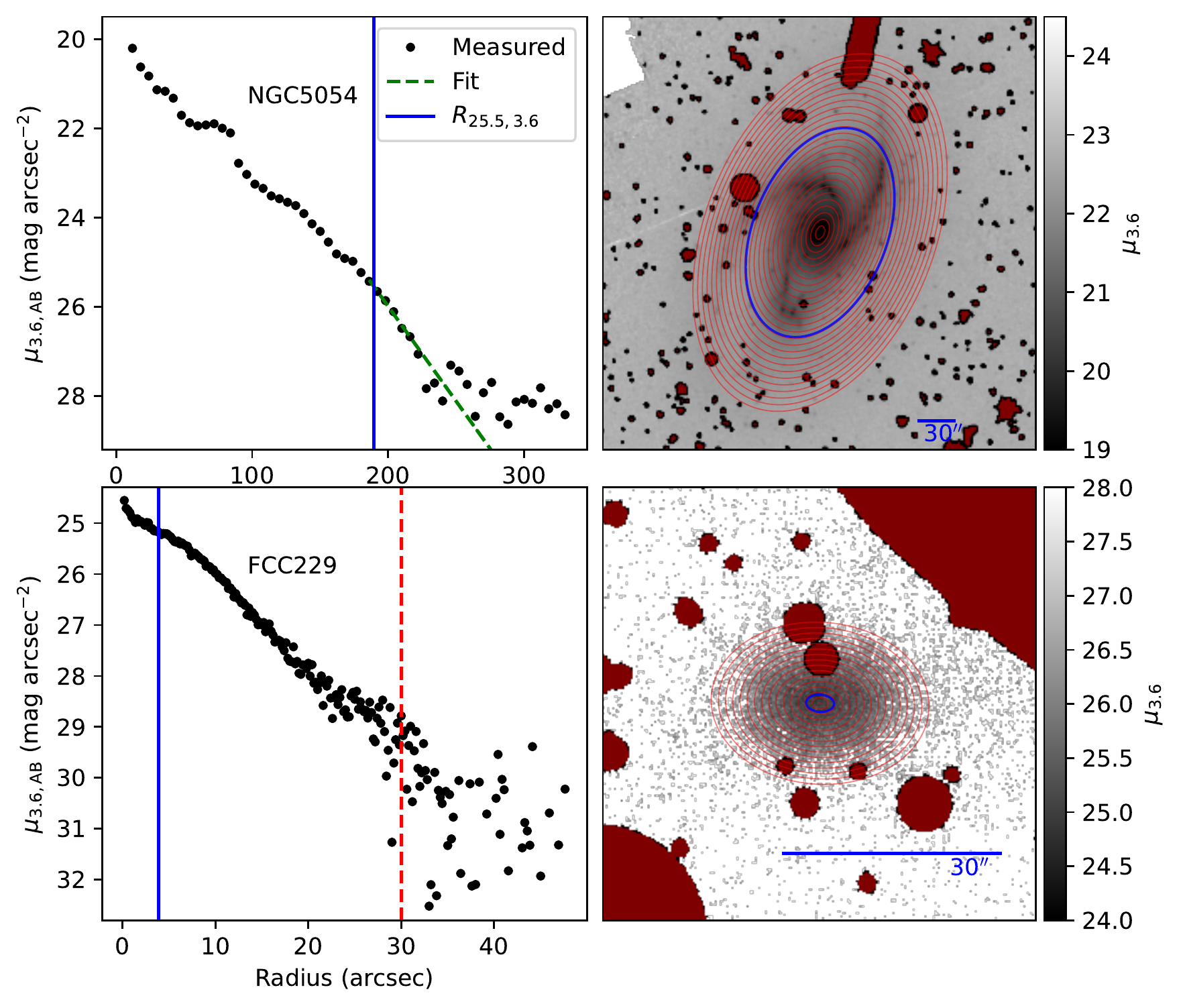}
    \caption{emph{Left} panels: example radial surface brightness profiles for an S$^{4}$G galaxy (NGC~5054, top row) and an FDS galaxy (FCC~229, or FDS11\textunderscore 0134, bottom row), deprojected to face-on surface brightness using the galaxies' measured axial ratios at $R_{25.5, 3.6}$.  For FCC~229 (an outlier from the IRSMR; see text), we show its mass surface density profile derived via Eq.~\ref{eq:mstar} converted to 3.6$\mu$m surface brightness via Eq.~\ref{eq:sigma36}.  Black points denote the measured profiles.  The green dashed line in the top-left panel shows a linear fit in the outskirts to smooth over uncertainties and project the profile beyond the image's noise limit; we did not apply such an extrapolation to the FDS galaxy profiles given the survey's greater depth.  The vertical blue lines show the locations of $R_{25.5, 3.6}$ for these deprojected profiles, which we use to derive our isomass radii.  \emph{Right} panels: 3.6$\mu$m image of NGC~5054 (top) and $r-$band image of FCC~229 (bottom), both with pixel values converted into 3.6$\mu$m surface brightness mimicking the left panels (color scales are shown on the right), with illustrative surface photometry annuli overplotted (for demonstration) as thin red ellipses (for NGC~5054, corresponding to every other point in the top left panel, and for FCC~229, corresponding to every tenth point in the bottom-left panel, truncated arbitrarily at 30\arcsec, denoted by the red vertical line in the bottom-left panel).  The blue ellipses show the locations of $R_{25.5, 3.6}$.  Blue horizontal lines are 30\arcsec in length, for scale.  Masked regions, which were ignored during surface photometry, are displayed in dark red.  In both panels, north is up and east is to the left.
    \label{fig:demo}}
\end{figure*}

To conduct our analysis, we require four parameters: total stellar mass, isomass radius, S\'{e}rsic index, and central mass surface density.  For the FDS sample, we had available integrated magnitudes, central surface brightnesses, and S\'{e}rsic indexes in $g^{\prime}$, $r^{\prime}$, and $i^{\prime}$.  Following \citet{venhola19}, we converted absolute magnitudes $M_{r^{\prime}}$ to stellar masses with the equations from \citet{taylor11}:
\begin{equation}
  \log \frac{\mathcal{M}_{*}}{\mathcal{M}_{\odot}} = 1.15 + 0.7 (g-i) +0.4 (r-i) -0.4 M_{r^{\prime}}.
\label{eq:mstar}
\end{equation}
Similarly, for surface density in units of $\mathcal{M}_{\odot}$~pc$^{-2}$:
\begin{eqnarray}
\log \Sigma_{*} &=& 9.78 +0.7\ (g-i) +0.4\ (r-i) -0.4 \mu_{r^{\prime}}.
\label{eq:sigma}
\end{eqnarray}
According to \citet{taylor11}, the accuracy of  Eq.~\ref{eq:mstar} is better than 0.1~dex.  Since the sample used by \citet{taylor11} did not contain dwarfs with $\log(\mathcal{M}_{*}/\mathcal{M}_{\odot})<7.5$, \citet[see their Fig.~2]{venhola19} compared this formula with an independent conversion based on \citet{bell01}, giving mass estimates consistent within 10\% for all FDS galaxies. We adopt the same systematic uncertainty in our study.  Flux profiles for FDS galaxies in $g^{\prime}$, $r^{\prime}$, and $i^{\prime}$ bands are from \citet{su21}, where they were sampled in elliptical annulae, with shapes and orientations corresponding to each galaxy's outer isophotes in the $r^{\prime}$-band images.  For typical $g-i=0.8, \ r-i =0.3$, $\mu_{r^{\prime}}=24, 25, 26, 27, 28$, and $29$ corresponds to $\Sigma{*} = 8, 3.2, 1.2, 0.5, 0.2$, and $0.08 \mathcal{M}_{\odot}$~pc$^{-2}$.

We assume all Fornax Cluster galaxies are at the same distance \citep[20~Mpc, following][]{blakeslee09, su21}, which incurs some uncertainty on their mass estimates.  Uncertainty on the Fornax Cluster distance itself merely moves mass-based scaling relations left and right, while the primary uncertainty influencing the widths of said relations is the location of each galaxy within the cluster.  Fornax's virial radius is 0.7~Mpc \citep{drinkwater01}, which suggests a maximum distance uncertainty on a particular Fornax galaxy of 1.4~Mpc.  Propagating this through Eq.~\ref{eq:mstar}, the maximum uncertainty on our derived Fornax Cluster member masses is 0.06~dex.  This assumes the galaxy in question is located in projection exactly in the Fornax Cluster centre, and that it lies, in truth, at one of the cluster edges (near or far side), and so is a worst-case scenario.  The true distance-based uncertainty will generally be smaller than this and depends on where the galaxy in question lies in projection with respect to the cluster centre (assuming the cluster is roughly spherical).  Thus, even in the worst case, the distance uncertainty on Fornax Cluster members' stellar masses is quite small ($\ll 0.06$).

For the S$^{4}$G and its ETG extension, stellar mass is converted to 3.6$\mu$m AB flux as follows:
\begin{eqnarray}
\log \frac{\mathcal{M}_{*}}{\mathcal{M}_{\odot}} &=& 2.12  -0.4 \ M_{3.6,\,AB}\\
\log \Sigma_{*} &=& 10.76 -0.4 \ \mu_{3.6,\,AB}.
\label{eq:sigma36}
\end{eqnarray}
The distance uncertainty on this population is much higher than for the Fornax members (typical distance uncertainties for S$^{4}$G galaxies are $\sim$3~Mpc), but because we are using this population primarily for a broad comparison between dwarfs and high-mass galaxies, this uncertainty does not impact any of our conclusions.  Combining Equations \ref{eq:sigma} and \ref{eq:sigma36}, we have the following relation between FDS colours and S$^{4}$G magnitudes:
\begin{eqnarray}
  \mu_{3.6,\,AB} &=& 2.45 -1.75 (g-i) - (r-i) + \mu_{r^{\prime}}.
  \label{eq:mu36_r}
\end{eqnarray}
For typical $g-i=0.8, \ r-i =0.3$ we have $\mu_{3.6,\,AB}\approx \mu_{r^{\prime}}+0.75$.

We experimented with different methods for calculating the colour terms in Eq.~(\ref{eq:sigma}). Since using the $g^{\prime}-i^{\prime}$ and $r^{\prime}-i^{\prime}$ profiles directly gives very noisy surface density profiles, we decided to use average colour differences calculated within the annulus extending from 0.25$R_{\rm eff}$ to 1.5$R_{\rm eff}$ in the $r^{\prime}$-band. The use of average colour should be a fair approximation, as cluster dwarfs tend not to have strong color gradients \citep[e.g.,][]{urich17}.  The chosen limits exclude the influence of a possible nucleus component, and also avoid the noisy outer parts.

From the surface density profiles, we constructed various isodensity radii, corresponding to either fixed $M_{3.6,\,AB}$ or fixed $\Sigma_{*}$ levels.  In practice, we recorded the innermost radius where density falls below the given isodensity level and made a linear fit to the profile around this radius using the 7 nearest profile points, to obtain a less noisy estimate of the distance where the density level is crossed.  To estimate the uncertainties of isodensity radii, we repeated the above procedure for profiles with $\Sigma_{*}(r) \pm \Delta \Sigma_{*}(r)$, where $\Delta \Sigma_{*}(r)$ is the uncertainty due to sky background subtraction. The resulting relative uncertainties in isophotal radii are typically from about 0.1~dex at $\mathcal{M}_{*}= 10^{6}\mathcal{M}_{\odot}$ to 0.01~dex at $\mathcal{M}_{*}= 10^{8}\mathcal{M}_{\odot}$. For comparison, if 'raw' colour profiles were used directly, the uncertainties would be roughly a factor 3 larger.

For the S$^{4}$G sample, all of the necessary parameters were already available through Pipelines 3 and 4.  Briefly, Pipeline 3 produced radial surface brightness profiles and curves of growth through annular and elliptical aperture photometry, with apertures increasing in semi-major axis using both 2\arcsec \ (approximately the IRAC FWHM) and 6\arcsec \ (for enhanced S/N in the galaxy outskirts) increments.  Integrated magnitudes in 3.6$\mu$m and 4.5$\mu$m were estimated using the radial profile of the slope of the curve of growth, extrapolated to $dm_{3.6,\,AB}/dr=0$ where the enclosed magnitude is by definition the total magnitude \citep[examples of this procedure can be found in][]{munozmateos15, watkins22}.  These magnitudes were then converted to stellar mass using either the relation given by \citet[for the original S$^{4}$G sample]{eskew12} or that given by \citet[for the ETG extension sample]{meidt12}, with negligible differences between relations.  Two isophotal radii, corresponding to the $\mu_{3.6,\,\rm AB} = 25.5$ and $26.5$ mag arcsec$^{-2}$ isophotes ($R_{25.5,\,3.6}$ and $R_{26.5,\,3.6}$), were derived using the 6\arcsec-width surface brightness profiles, interpolated to enhance the precision.

For both FDS and S$^{4}$G, to account for the galaxies' elliptical projected shapes, we multiplied our density profiles with the axial ratios ($b/a$) of the elliptical annuli \citep[following][]{kent85}, using only those of outer radii (thereby applying the same correction factor to every isophote in a galaxy).  Interestingly, alternative deprojection strategies, which nominally take into account oblate or prolate shapes (useful mainly for the ETG sample) resulted in higher scatter in our scaling relations compared to the simple axial ratio multiplication.  Why this is the case is not clear; we will explore this in more detail in a companion paper (Salo et al. in prep.).

The noise in the profiles' outskirts for the S$^{4}$G galaxies (below $\mu_{3.6} \approx 25$) was often quite high, even exceeding $\mu_{3.6,\, \rm AB} = 25.5$ in some cases.  To address this, we re-derived the S$^{4}$G isophotal radii using linear fits (with limits chosen by eye) to the noisy parts of each profile and estimated the radii from these fits instead of the measured profiles, extrapolating beyond the noise limits when necessary.  We show an example of this process in Fig.~\ref{fig:demo}.  We note that we did not perform such an extrapolation for the FDS sample galaxies, as the much greater depth ($3$--$4$ mag arcsec$^{-2}$) of these images made it unnecessary.

Central mass surface densities and S\'{e}rsic indexes for both FDS and S$^{4}$G galaxies came from GALFIT decompositions, via the IDL interface GALFIDL \citep{salo15}.  The full description of this process can be found in Sec.~4.2 of \citet{su21} for the FDS galaxies, and in Sec.~2 of \citet{salo15} for the S$^{4}$G galaxies.  We used single-component S\'{e}rsic fits for both the FDS and S$^{4}$G to derive S\'{e}rsic components, to maintain consistency across surveys \citep[only single-component fits are currently available for the S$^{4}$G ETG extension;][]{watkins22}.  The nucleus component, if present, was accounted for by an unresolved central source modeled with the image PSF; thus, its influence on S\'{e}rsic parameters was removed.  By contrast, we derived our stellar masses and isomass radii through aperture photometry, using the methods and conversion formulas described above.

For the remainder of this paper, we work in units of stellar mass ($\mathcal{M}_{*}$) and stellar mass surface density ($\Sigma_{*}$).  We denote the central stellar mass surface density $\Sigma_{*}(R=0)$ as $\Sigma_{*,\,0}$.  As a balance between radial extent and S/N across both surveys, we use as our fiducial isophotal radius $R_{25.5,\,3.6}$, which corresponds to an isomass radius of $R_{3.63\, \mathcal{M}_{\odot}\,\rm{pc}^{-2}}$ (hereafter, $R_{3.63}$).  This radius corresponds to the outskirts of massive galaxies, but is located close to the centre for low-mass galaxies given their lower surface brightnesses.  We discuss the implications of this choice throughout.

\section{Results}\label{sec:results}

\subsection{Scaling relations}\label{sec:scaling}

\begin{figure*}
    \centering
    \includegraphics[scale=1.0]{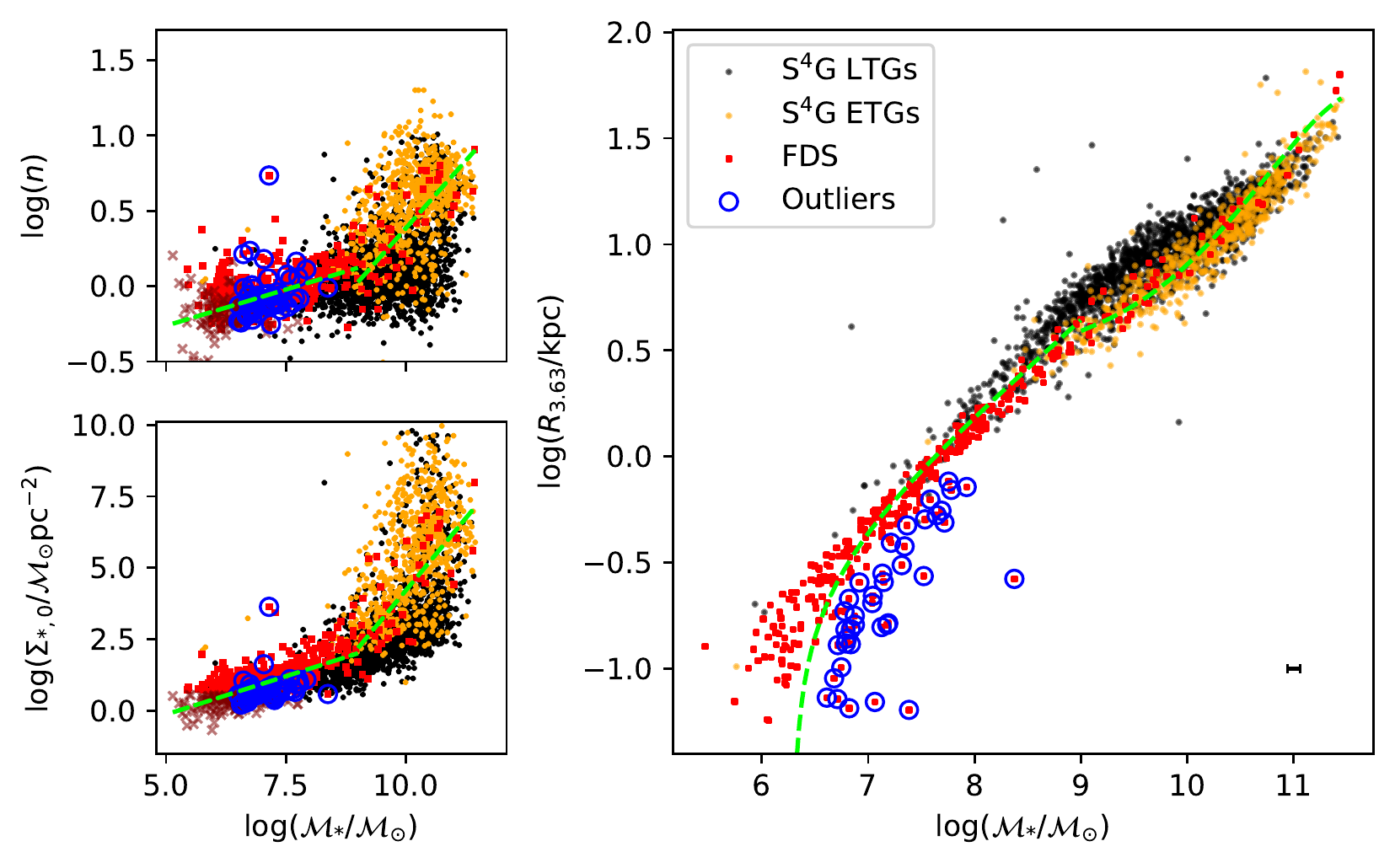}
    \caption{Scaling relations for our galaxy sample.  The \emph{top-left} panel shows S\'{e}rsic index vs. stellar mass, in solar mass units.  S$^{4}$G galaxies are shown as black (LTG) and orange (ETG) points, while FDS galaxies are shown as red squares and x's.  Red x's denote galaxies with best-fit $\Sigma_{*,\,0} < 3.6$ (159 dwarfs total; see text).  The \emph{bottom-left} panel echoes the top-left, but shows stellar mass against central mass surface density.  In both panels, the dashed green lines show log-linear fits to the relations, separated at $\log(\mathcal{M}_{*}/\mathcal{M}_{\odot})=9.0$.  The \emph{right} panel shows the relation between the $\Sigma_{*} = 3.63 \mathcal{M}_{\odot}$~pc$^{-2}$ isomass radius $R_{3.63}$ and stellar mass (the IRSMR).  The colour scheme is the same as in the left two panels  The dashed green line shows the $\log(R_{3.63})$--$\log(\mathcal{M}_{*}/\mathcal{M}_{\odot})$ relation predicted by the log-linear fits in the left two panels.  Points circled in blue in all panels are the outliers from the IRSMR, which fall below the relation by more than the root-mean-square of the residuals from the predicted relation.  Galaxies denoted by red x's in the left two panels are excluded in this panel.  The black errorbar in the lower-right corner shows an estimate of the maximum possible mass uncertainty on Fornax Cluster galaxies based on the cluster's virial radius (0.06~dex; see text), though the true distance-based uncertainty for a given galaxy is much smaller than this.
    \label{fig:main}}
\end{figure*}

\begin{figure*}
    \centering
    \includegraphics[scale=1.0]{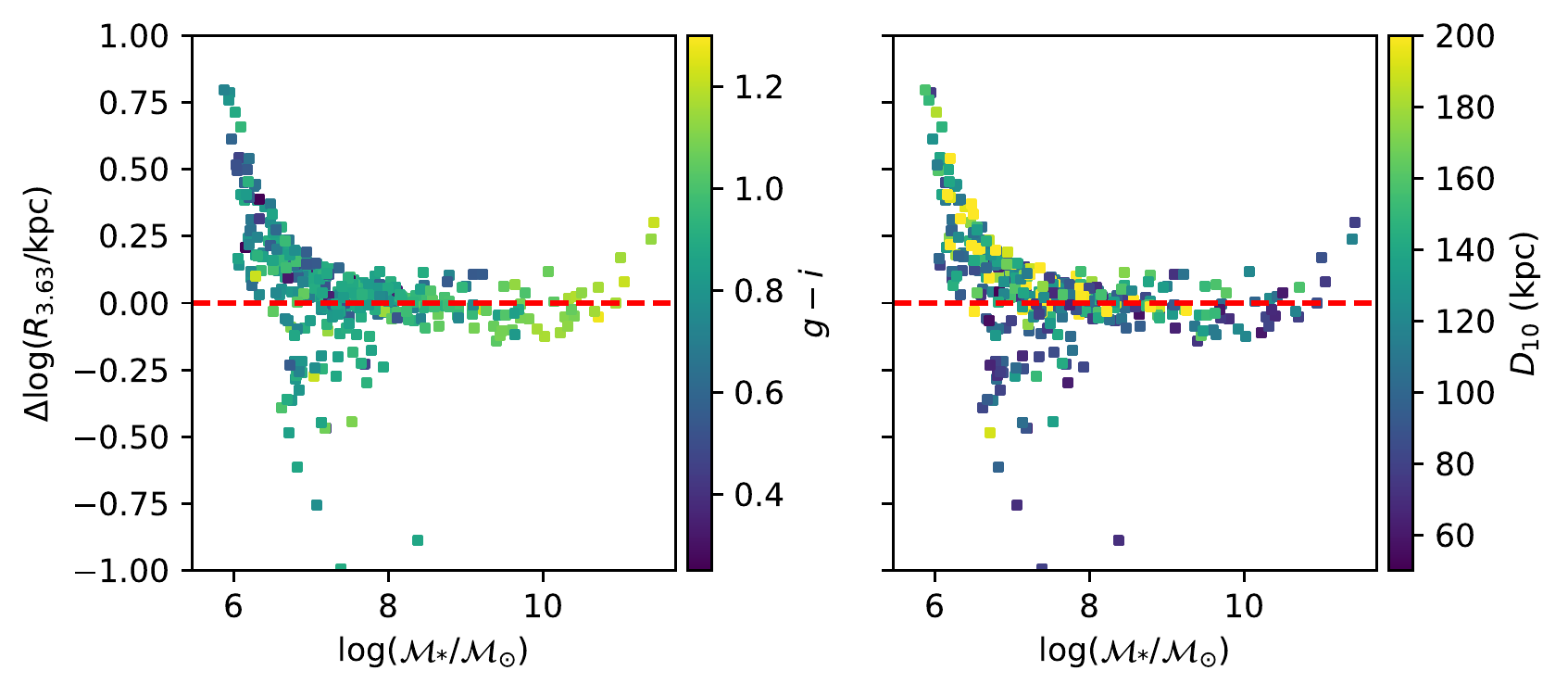}
    \caption{Residuals from our best-fit isomass-radius--stellar-mass relation for FDS galaxies, colour-coded by two different variables: $g^{\prime}-i^{\prime}$ colour (\emph{left} panel), and projected distance to each galaxy's 10$^{\rm th}$-nearest neighbour (\emph{right} panel).  The dashed red lines show zero, i.e. the location of the predicted IRSMR (the green dashed curve in the right panel of Fig.~\ref{fig:main}) in this space.
    \label{fig:colored}}
\end{figure*}

\begin{table}
\centering
\caption{Log-linear fit parameters for our scaling relations, of the form $Y = k\log(\mathcal{M}_{*}/\mathcal{M}_{\odot}) + c$.  Uncertainties are the standard errors on each fit parameter.  We also provide the reduced $\chi^{2}$ of the residuals of each fit.
\label{tab:fits}}
\begin{tabular}{llccc}
\hline\hline
Mass range & $Y$ & k & c & $\chi^{2}_{\nu}$ \\
\hline \\
$\log(\mathcal{M}_{*}/\mathcal{M}_{\odot})\leq9.0$ & $\log(n)$ & 0.10$\pm$0.01 & -0.74$\pm$0.05 & 0.018 \\
$\log(\mathcal{M}_{*}/\mathcal{M}_{\odot})\leq9.0$ & $\log(\Sigma_{*,\,0})$ & 0.54$\pm$0.03 & -2.81$\pm$0.18 & 0.211 \vspace{2pt} \\
$\log(\mathcal{M}_{*}/\mathcal{M}_{\odot})>9.0$ & $\log(n)$ & 0.36$\pm$0.04 & -3.20$\pm$0.42 & 0.089 \\
$\log(\mathcal{M}_{*}/\mathcal{M}_{\odot})>9.0$ & $\log(\Sigma_{*,\,0})$ & 2.02$\pm$0.24 & -16.00$\pm$2.39 & 2.903 \\
\hline
\end{tabular}
\end{table}

We present the FDS and S$^{4}$G scaling relations in Fig.~\ref{fig:main}.  In the left two panels, we show the relations between stellar mass $\mathcal{M}_{*}$ and S\'{e}rsic index $n$ (top left) and between $\mathcal{M}_{*}$ and central mass surface density $\Sigma_{*,\,0}$ (bottom left).  We show S$^{4}$G late-type galaxies (LTGs) in black (defined as having morphological $T-$types $> 0$) and S$^{4}$G ETGs (defined as having morphological $T-$types $\leq 0$) in orange. We show FDS galaxies as red squares and x's.  Galaxies marked with red x's denote those with best-fit $\Sigma_{*,\,0} < 3.63 \mathcal{M}_{\odot}$~pc$^{-2}$, which ostensibly have no isomass radius $R_{3.63}$.  Green dashed lines show log-linear fits to these relations, made using Scikit-learn's \citep{scikit} Huber Regressor \citep{huber11} Python package, with three iterations of $3\sigma$ rejection.  We provide the parameters of these fits in Table~\ref{tab:fits}, alongside values of the reduced $\chi^{2}$ of each fit, defined as the $\chi^{2}$ statistic of the residuals normalized by the number of degrees of freedom.

We fit two regression lines to each relation, divided at $\log(\mathcal{M}_{*}/\mathcal{M}_{\odot})=9.0$, as the shapes of the relations imply a dichotomy between low-mass and high-mass galaxies.  \citet{derijcke09} noted a similar dichotomy in a sample combined from many different surveys, and \citet{calderon15} found hints of this division in the Antlia Cluster, made clearer by including data from other clusters and groups \citep[see also:][]{kormendy12}.  While the high-mass galaxy sample in Fornax is sparse, it clearly follows the same relation as the much larger massive ETG sample from the S$^{4}$G, which very clearly has a distinct slope compared to the Fornax dwarf population even considering the high-mass relation's large scatter.  Therefore, with the benefit of the large sample sizes and consistent photometric methodologies between the surveys, we conclude that this dichotomy is real.  The most appropriate mass division is less clear, as it remains unclear why such a dichotomy might be present.  We discuss this further in Section~\ref{sec:conclusions}, but we find that any choice between $\log(\mathcal{M}_{*}/\mathcal{M}_{\odot})=8.5$--$9.5$ does not alter our conclusions appreciably.

Though the fits at the high-mass end appear slightly skewed, alternative methods produce very similar regression lines, including Bayesian Monte Carlo regression.  Inverting the axes before fitting produces fits biased in the opposite direction.  This slight skewness is likely a result of the high scatter, which tends to flatten predicted slopes in linear regression models, and the near-vertical correlation.  The latter may partly arise from systematic errors in the fitting: \citet{watkins22} showed that for many high-mass ETGs, single-component fits (whether GALFIT decompositions or S\'{e}rsic fits to radial profiles) tend to over-estimate S\'{e}rsic index and central surface brightness, leading to an excess of high values that may skew the measured scaling relation.  While more robust estimates of $n$ and $\Sigma_{*,\,0}$ (for example, targeting specific regions of the galaxies via multi-component fitting, or using higher resolution imaging data) might improve this scatter \citep[e.g.,][and references therein]{graham19}, our focus in this paper is primarily on the low-mass galaxies, which are well-fit by single-component profiles \citep{venhola19, su21} and which, perhaps consequently, show significantly less scatter in their corresponding relations.

Likewise, for the dwarf population, we have assumed these follow scaling relations.  However, \citet{derijcke09} found that dwarfs in their sample show no power law relation between $n$ and $\mathcal{M}_{*}$, merely scattering between values of $0.5 < n < 1.0$.  If we instead assume a constant value for $n$ of 0.85 (the median value of $n$ among the FDS dwarf population), we find an RMS value of 0.151, slightly higher than the best-fit power-law relation value of 0.134 (Table~\ref{tab:fits}).  This would seem to argue that this relation is indeed better described by a power law, although the difference is subtle enough that an assumed constant value is not unreasonable.  We discuss the implications of this choice in more detail in Sec.~\ref{sec:scatter}.  Regarding $\Sigma_{*,\,0}$, however, a constant value fit increases the RMS from 1.147 to 1.606, a large enough difference that the power-law fit seems justified.

The right panel of Fig. \ref{fig:main} shows the relationship between $\mathcal{M}_{*}$ and $R_{3.63}$, the IRSMR.  The colour scheme for S$^{4}$G galaxies is the same as the left two panels.  The green dashed line shows the predicted $\log(R_{3.63})$--$\log(\mathcal{M}_{*})$ relationship derived from the log-linear fits in the left two panels.  Points circled in blue are outliers from this relation, which we define as points falling below the curve by more than one RMS of the residuals of FDS galaxies about this relation (RMS$=0.170 \log({\rm kpc})$).  We discuss these outliers in more detail in the following section.

The curve's shape follows from the S\'{e}rsic function.  First, we assume that the relation between $\mathcal{M}_{*}$ and $n$ has the form
\begin{equation}\label{eq:mstar_n}
    \log(n) = k_{n}\log(\mathcal{M}_{*}) + c_{n},
\end{equation}
where $k_{n}$ and $c_{n}$ are constants, and that the relation between $\mathcal{M}_{*}$ and central surface brightness has the form
\begin{equation}\label{eq:mstar_mu0}
    \mu_{0} = k_{\mu}\log(\mathcal{M}_{*}) + c_{\mu},
\end{equation}
where $k_{\mu}$ and $c_{\mu}$ are two additional constants.  The S\'{e}rsic profile, when expressed in units of surface brightness, has the following form:
\begin{equation}\label{eq:sersic1}
    \mu(r) = \mu_{0} + \frac{2.5b_{n}}{\ln10}\left(\frac{r}{R_{\rm eff}}\right)^{1/n}.
\end{equation}
The constant $b_{n}$ is defined such that 
\begin{equation}\label{eq:bn}
    \Gamma(2n) = 2\gamma(2n, b_{n})
\end{equation}
\citep{ciotti91}, where $\Gamma$ and $\gamma$ are the complete and incomplete gamma functions, respectively \citep[for a summary of the properties of the S\'{e}rsic function, see][]{graham05}.  Likewise, if we define a term $A$ such that
\begin{equation}\label{eq:a}
    A = -2.5\log\left(2\pi n \cdot \frac{e^{b_{n}}}{b_{n}^{2n}}\right)\Gamma(2n)
\end{equation}
$R_{\rm eff}$ can be rewritten in terms of the total magnitude $m_{\rm tot}$ (or, equivalently, the total mass) of the profile as follows:
\begin{equation}\label{eq:reff}
    \log(R_{\rm eff}) = 0.2\left(\mu_{0} + \frac{2.5b_{n}}{\ln10} + A - m_{\rm tot}\right)
\end{equation}
\citep[e.g., Eq.~5 of][]{graham05}.  An isophotal radius $R_{x}$ is defined such that $\mu(R_{x})=x$ mag arcsec$^{-2}$, so we can solve for this radius in Eq.~\ref{eq:sersic1}:
\begin{equation}\label{eq:riso_mstar}
    R_{x} = R_{\rm eff} \left((\mu(R_{x}) - \mu_{0}) \cdot \frac{\ln(10)}{2.5b_{n}}\right)^{n}.
\end{equation}
Substituting the relations defined in Eqs.~\ref{eq:mstar_n}, \ref{eq:mstar_mu0}, and \ref{eq:reff}, this becomes a relation purely between $R_{x}$ and $m_{\rm tot}$.  This is then directly convertible to an IRSMR for a specific choice of isophotal radius $\mu(R_{x})$.

The curvature of this predicted relation, shown in green in the right panel of Fig.~\ref{fig:main}, matches that in the data quite well.  There is a sharp downturn in the predicted relation at the lowest masses (Fig.~\ref{fig:main}, right-hand panels), resulting from the choice of isomass radius: the value of $\Sigma_{*,\,0}$ predicted by the log-linear fit in the low-mass regime reaches $3.6 \mathcal{M}_{\odot}$~pc$^{-2}$ at a stellar mass of $\log(\mathcal{M}_{*}/\mathcal{M}_{\odot}) \sim 6.3$, resulting in unphysical estimates of $R_{x}=R_{3.63}$.  This downturn is less severe when using larger isomass radii \citep[see Sec.~4.3, Fig.~7 from][]{graham19}, and its location is also sensitive to the log-linear fits from which the curve is derived (Eq.~\ref{eq:mstar_n} and \ref{eq:mstar_mu0}), making identification of IRSMR outliers difficult in the low-mass regime.

The predicted curve fits the ETGs much more closely than the LTGs at the high-mass end, but this is because the ETGs drive the two scaling relations used to derive the curve.  Most LTGs have best-fit $n=1$ regardless of mass when single-component models are used \citep{salo15}.

As noted by \citet{sanchezalmeida20}, the slope of the IRSMR depends on $n$, which in turn depends on $\mathcal{M}_{*}$, resulting in this gradually changing curvature.  While the values of $n$ and $\Sigma_{*,\,0}$ in our sample were derived through single-component GALFIT decompositions, the isophotal radii and total stellar masses are empirical (Sec.~\ref{sec:methods}), and so are independent of any assumptions about the light profiles' true shapes.  That the curvature in the relation predicted from the S\'{e}rsic decompositions matches the empirically derived relation so closely provides further evidence that the mass division in these relations is real, and that there are real structural differences between galaxies with masses greater and less than $\log(\mathcal{M}_{*}/\mathcal{M}_{\odot}) \sim 9$.

In Fig.~\ref{fig:colored}, we show the residuals from our fiducial IRSMR for FDS galaxies, colour-coded by two different properties: integrated $g^{\prime} - i^{\prime}$ colour (eft panel), and projected distance to the tenth nearest-neighbour, in kpc \citep[$D_{10}$, right panel, a measure of local density;][]{dressler80}.  Galaxies on and off the IRSMR have similar colours, however outlier galaxies falling below the relation seem to be generally found in denser local environments than those on or above the relation, suggesting they could be a distinct population.  Also, the bottom-left panel of Fig.~\ref{fig:main} shows that the outliers are frequently low in $\Sigma_{*,\,0}$, mostly falling below the best-fit line.  Hence, these merit further examination.  However, defining outliers from the IRSMR, particularly at the low-mass end, is not trivial, particularly given how the uncertainties on the $\mathcal{M}_{*}$--$n$ and $\mathcal{M}_{*}$--$\Sigma_{*,\,0}$ relations propagate into uncertainties on the IRSMR.  Therefore, before we can define outliers, we must discuss the uncertainty on the IRSMR in more detail.

\subsection{On the IRSMR scatter and theoretical shape}\label{sec:scatter}

\begin{figure*}
    \centering
    \includegraphics[scale=1.0]{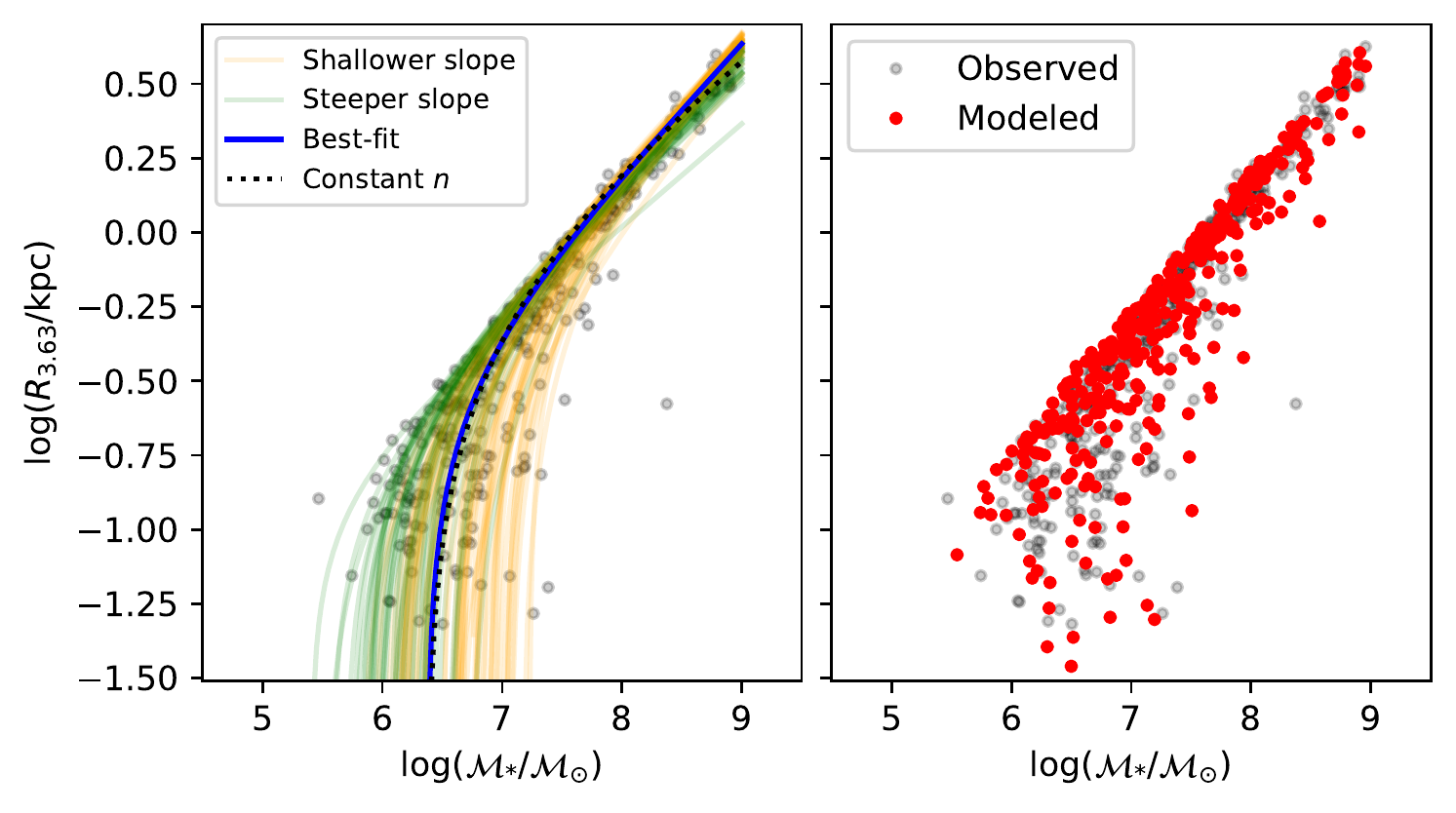}
    \caption{Tests demonstrating the nature of the theoretical IRSMR.  The \emph{left} panel shows our fiducial IRSMR, for galaxies with masses below $\mathcal{M}_{*} < 10^{9} \mathcal{M}_{\odot}$.  We show our best-fit curve predicted from our log-linear fits to the $\mathcal{M}_{*}$--$n$ and $\mathcal{M}_{*}$--$\Sigma_{*,\,0}$ relations (Table~\ref{tab:fits}) in blue.  Orange and green lines show predicted curves in this space made by adjusting our best-fit slopes and intercepts by the error on the fit parameters in Table~\ref{tab:fits}, demonstrating the sensitivity of the curve's shape to the parameters.  Green lines have steeper slopes in the $\mathcal{M}_{*}$--$\Sigma_{*,\,0}$ relation, while orange lines have shallower slopes in this relation.  For comparison, we show the curve predicted by assuming a constant value of $n$ as a dotted black line, which is almost indistinguishable from our best-fit relation.  We plot the observed IRSMR for FDS dwarfs as black points underneath.  The \emph{right} panel shows the observed IRSMR for FDS dwarfs in black, while the red points show a theoretical IRSMR made by adding artificial scatter about the best-fit $\mathcal{M}_{*}$--$n$ and $\mathcal{M}_{*}$--$\Sigma_{*,\,0}$ relations before deriving the isomass radii from them.  The scatter here is preferentially downward, just as it is in the observed relation.
    \label{fig:theor}}
\end{figure*}

We show how the estimated slopes of the $\mathcal{M}_{*}$--$n$ and $\mathcal{M}_{*}$--$\Sigma_{*,\,0}$ relations affect the predicted shape of the IRSMR in the left panel of Fig.~\ref{fig:theor}.  Here, we show our best-fit IRSMR as a blue line, plotted over the observed FDS dwarf relation (faded black points), as well as 100 alternative predicted IRSMRs derived by adjusting the best-fit slopes and intercepts of the $\mathcal{M}_{*}$--$n$ and $\mathcal{M}_{*}$--$\Sigma_{*,\,0}$ relations for dwarf galaxies (Table~\ref{tab:fits}) by the standard error of the best-fit values, shown as green and orange lines.  For each line, we add to each fit parameter a value drawn from a normal distribution with zero mean and standard deviation equal to the standard error of that fit parameter.  Green lines here were derived using steeper slopes than our best-fit relation in the $\mathcal{M}_{*}$--$\Sigma_{*,\,0}$ relation, while orange lines were derived with shallower slopes (the spread about the best-fit curve displayed by each set of coloured curves arises from random changes to the y-intercepts in this relation and from changes to the $\mathcal{M}_{*}$--$n$ relation parameters).  Despite such small adjustments, these curves diverge significantly, with downturns occurring anywhere between $5.5 < \log(\mathcal{M}_{*}/\mathcal{M}_{\odot}) < 7$ for the majority of the perturbations, a far greater uncertainty than either photometric or distance-based uncertainty.  Therefore, a conservative definition of outliers would use only dwarfs with stellar masses above $\mathcal{M}_{*} \gtrsim 10^{7} \mathcal{M}_{\odot}$.

Additionally, to investigate the impact of our assumption of a power law relation between $n$ and $\mathcal{M}_{*}$ (Sec.~\ref{sec:scaling}), we also show the curve predicted by assuming that $n$ is constant among the FDS dwarfs as a black dotted line.  Here we have set $n$ to the median value of the whole population (0.85).  The black curve is almost indistinguishable from our best-fit curve; this choice, therefore, does not affect our selection of IRSMR outliers.

The scatter in the IRSMR is also driven by the scatter in the $\mathcal{M}_{*}$--$n$ and $\mathcal{M}_{*}$--$\Sigma_{*,\,0}$ relations.  We demonstrate this in the right panel of Fig.~\ref{fig:theor}, which shows our fiducial dwarf-galaxy IRSMR again, with the observed relation in faded black points and a simulated relation in red.  We produced these simulated values by adding scatter around our best-fit $\mathcal{M}_{*}$--$n$ and $\mathcal{M}_{*}$--$\Sigma_{*,\,0}$ relations before deriving predicted isomass radii from said relations, to investigate how the uncertainty on both $n$ and $\Sigma_{*,\,0}$ estimates affect the shape of the IRSMR.  Most of the scatter in these simulated values trends downward, just as we see in the observed relation.  This results partly from the use of logarithmic scaling in all of our relations, making the galaxies with properties at the low-$\sigma$ edges of their distributions stand out very clearly in the IRSMR.  For the dwarf population, which shows a small linear spread in $n$ (with values ranging between $\sim0.5$--$1$) compared to its linear spread in $\Sigma_{*,\,0}$ (with values ranging between $\sim1$--$100 \mathcal{M}_{\odot}$~pc$^{-2}$), this property makes the IRSMR quite useful for isolating the lowest surface-brightness objects in the sample.

The strongest effect of uncertainty therefore lies at the low-mass end, where the derived shape of the IRSMR is the most sensitive to the fit parameters.  With that constraint in hand, we can compare the physical properties of IRSMR outliers to those of dwarfs lying close to the relation.  Before we can sensibly interpret any populations differences, however, we must first disentangle any inter-correlations among the galaxies' integrated parameters.

\subsection{Principal component analysis of integrated parameters}\label{sec:pca}

\begin{table}
\centering
\caption{Principal component analysis coefficients for three principal components, using all galaxies from the Fornax Cluster sample.
\label{tab:pca}}
\begin{tabular}{lccc}
\hline\hline
Parameter & PC1 & PC2 & PC3 \\
\hline \\
$\log(R_{3.63})$ & 0.458 & 0.083 & 0.161 \\
$\log(R_{\rm eff})$ & 0.411 & -0.134 & 0.503 \\
$\log(\mathcal{M}_{*}/\mathcal{M}_{\odot})$ & 0.481 & 0.018 & 0.164 \\
$n$ & 0.408 & 0.055 & -0.311 \\
$\log(\Sigma_{*,\,0})$ & 0.452 & 0.127 & -0.320 \\
$D_{\rm cluster}$ & -0.090 & 0.560 & 0.388 \\
$D_{10}$ & -0.086 & 0.598 & 0.279 \\
$\log(\textrm{PI})$ & -0.056 & -0.533 & 0.516 \\
\hline
\end{tabular}
\end{table}

The integrated quantities we are using to compare the outliers and IRSMR dwarfs are, unfortunately, inter-correlated.  Therefore, to investigate which of these quantities are the most important for driving any differences we find between populations, we performed a principal component analysis \citep[PCA;][]{pearson01} using all of the structural and environmental parameters we have available (excluding colour, which we found adds very little information while increasing the complexity of the PCA results).  PCA is a means of reducing the dimensionality of a dataset by transforming the dataset to a new, orthogonal coordinate system based on the normalized covariance matrix of the data.  Each principal component is a representation of this covariance matrix, such that the first component represents the axis along which the dataset's variance is highest, the second component the axis along which the variance is second-highest, and so on, with each component mapped to the dataset's parameters by a vector of weights.  For the Fornax Cluster galaxies only, we compare the following parameters: isomass radius ($R_{3.63}$), half-light radius ($R_{\rm eff}$), stellar mass ($\mathcal{M}_{*}/\mathcal{M}_{\odot}$), S\'{e}rsic index ($n$), central mass surface density ($\Sigma_{0,\,*}$), projected 10$^{\rm th}$-nearest neighbour distance, projected cluster-centric distance in degrees ($D_{\rm cluster}$), and a dimensionless perturbation index (PI).  PI is defined, for a galaxy $j$, as
\begin{equation}\label{eq:pi}
    PI_{j} = \sum_{i} \left(\frac{\mathcal{M}_{i}}{\mathcal{M}_{j}}\right)\left(\frac{R_{\rm eff,j}}{D_{i}}\right)^{3}
\end{equation}
where $\mathcal{M}_{i}$ is the stellar mass of the $i^{\rm th}$ neighbouring galaxy, $R_{\rm eff,\,j}$ is the $j^{\rm th}$ galaxy's effective radius, and $D_{i}$ is the separation (projected, in our case) between the $j^{\rm th}$ galaxy and the $i^{\rm th}$ galaxy \citep[see Eq.~3 of][]{jackson21}.  This is a measure of the total tidal influence from a given galaxies' neighbours in a system, first utilized by \citet{byrd90}.  It is akin to the earlier Dahari parameter \citep{dahari84}, but uses the galaxies' effective radii rather than their optical diameters.

We show these weights for the first three principal components of Fornax galaxies in Table~\ref{tab:pca}, which explain 51\%, 27\%, and 10\% of the dataset's variance, respectively.  The absolute value of these weights represents the strengths of the correlations between parameters, and the sign shows whether these are correlations or anti-correlations.

The first principal component includes contributions mainly from the structural parameters (size, $\mathcal{M}_{*}$, $n$, and $\Sigma_{*,\,0}$), all of which are positively correlated.  The second component includes primarily the environmental parameters ($D_{\rm cluster}$, $D_{10}$, and PI, which is anti-correlated with the others, as expected).  The third component is a mixture, but the strongest weights here apply to $R_{\rm eff}$ and PI, which show a positive correlation.  This is expected given that PI is defined using the cube of $R_{\rm eff}$ for a given galaxy (Eq.~\ref{eq:pi}), but given that this correlation emerges only in the third component (which explains only $\sim 10$\% of the data's total variance), it appears that PI remains predominantly a measure of environment for the FDS sample.  Nonetheless, this strong correlation with $R_{\rm eff}$ makes PI alone a biased metric of tidal perturbation strength among cluster galaxies; we performed an experiment by randomly rearranging our dwarfs radially within the Fornax Cluster, and found that regardless of radial position, dwarfs with large $R_{\rm eff}$ always show the highest values of PI.  Tidal perturbation indices, therefore, should be used primarily for interpretation of galaxy pair interactions, as they were originally designed \citep{dahari84}, and not for a measure of tidal strength within a large, bound system like a galaxy cluster.  We therefore favour $D_{10}$ (which explains most of the variance in the second principal component) and $D_{\rm cluster}$ as our primary probes of environment.

Having now established that intrinsic parameters and environmental parameters appear separable in terms of their contributions to the sample's total variance, interpretation of population comparisons is straight-forward.  We show such comparisons in the following section.

\subsection{Outliers from the IRSMR}\label{sec:outliers}

\begin{table*}
\centering
\caption{Parameter comparisons for FDS dwarfs on and off the IRSMR.  The sample sizes for the populations on and off the relation are 246 and 48, respectively, with median masses of $\log(\mathcal{M}_{*}/\mathcal{M}_{\odot}) = 7.5$ and $7.1$. (1)$-$ parameter name; (2)$-$ parameter unit; (3)$-$ median of parameter for galaxies on IRSM; (4)$-$ bootstrapped error on (3); (5)$-$ median of parameter for outlier galaxies; (6)$-$ bootstrapped error on (5); (7)$-$ Mann-Whitney U test statistic comparing both populations; (8)$-$ $p-$value associated with (7)
\label{tab:comp}}
\begin{tabular}{lccccccccc}
\hline\hline
Parameter & Unit & Med$_{\rm on}$ & $\sigma_{\rm on}$ & Med$_{\rm off}$ & $\sigma_{\rm off}$ & $U$ & $p_{U}$\\
\hline \\
$g^{\prime}-i^{\prime}$ & AB mag & 0.875 & 0.006 & 0.877 & 0.012 & 4552 & 0.917 \\
$r^{\prime}-i^{\prime}$ & AB mag & 0.271 & 0.006 & 0.259 & 0.018 & 5005 & 0.268 \\
$R_{\rm eff}$ & kpc & 0.842 & 0.041 & 0.924 & 0.062 & 4012 & 0.275 \\
$n$ & -- & 0.970 & 0.026 & 0.834 & 0.038 & 5714 & 7.354e-03 \\
$\Sigma_{*,\,0}$ & $\mathcal{M}_{\odot}$~pc$^{-2}$ & 1.299 & 0.026 & 0.714 & 0.026 & 8127 & 9.746e-16 \\
$D_{\rm cluster}$ & deg & 1.574 & 0.115 & 1.026 & 0.106 & 6051 & 6.092e-04 \\
$D_{10}$ & kpc & 120.113 & 4.747 & 88.696 & 5.387 & 6499 & 9.837e-06 \\
$\log(\rm{PI})$ & -- & -3.181 & 0.072 & -2.158 & 0.147 & 2100 & 9.838e-08 \\
\hline
\end{tabular}
\end{table*}

As discussed by \citet{watkins22}, all of the prominent outliers from the IRSMR at high stellar mass are edge-on galaxies, with axial ratios artificially increased by thick disks or stellar halos; the use of the \citet{kent85} method for isophotal circularization is thus not an adequate deprojection for such galaxies, leading to skewed isomass radius estimates.  This is not the case for the low-mass ($\mathcal{M}_{*} \leq 10^{9} \mathcal{M}_{\odot}$) outliers, which have a random assortment of axial ratios, suggesting that their outlier status could be physical in nature.

Fig.~\ref{fig:colored} shows evidence that these low-mass outliers are somehow distinct, such that outliers below the relation seem to exist in notably denser local environments than those on the relation.  Therefore, we examined the bulk properties of these outliers using a variety of different intrinsic and environmental parameters.  Given the brighter surface brightness limit and lower resolution of the S$^{4}$G imaging, we exclude S$^{4}$G galaxies from this low-mass sample for the remainder of our analysis.  We also exclude all FDS galaxies with central surface mass densities $\Sigma_{*,\,0} < 3.63 \mathcal{M}_{\odot}$~pc$^{-2}$.

Table~\ref{tab:comp} shows these parameter comparisons between dwarf galaxy populations lying on the IRSMR and outliers from it, including all parameters we investigated in Table~\ref{tab:pca}, as well as integrated $g^{\prime}-i^{\prime}$ and $r^{\prime}-i^{\prime}$ colours.  As stated in Sec.~\ref{sec:methods}, we define outliers as galaxies falling more than one RMS below the predicted relation.  Anything within one RMS we define as being on the relation, excluding dwarfs with $\Sigma_{*,\, 0} < 3.63 \mathcal{M}_{\odot}$~pc$^{-2}$ (240 galaxies).  We also exclude galaxies with masses $\log(\mathcal{M}_{*}/\mathcal{M}_{\odot}) \leq 6.5$, to avoid the sharp downturn in the predicted curve (Sec.~\ref{sec:scaling}).  We do not include outliers which fall above the relation because the scatter in the relation is preferentially toward low isomass radii (Sec.~\ref{sec:scatter}), making high-isomass radius outliers difficult to identify using this relation.  Adjusting our outlier definition to exclude galaxies with $\mathcal{M}_{*} < 10^{7} \mathcal{M}_{\odot}$ (Sec.~\ref{sec:scatter}) and to accept only galaxies $>2\times$RMS from the relation does not appreciably alter our results, though the outlier sample size decreases from 52 to only 14.  The uncertainty on the shape of the theoretical curve at the low-mass end dominates the outlier selection over the uncertainty on $R_{3.63}$ and $\log(\mathcal{M}_{*}/\mathcal{M}_{\odot})$, and hence our statistical tests are also robust to these uncertainties.  We discuss the photometric and calibration uncertainties in more detail in Sec.~\ref{sec:app_errs}.

For each population and each parameter, we estimated the median values and the bootstrapped errors on those medians, derived as the standard deviation of the median values of $N=1000$ randomly chosen sub-samples of each parameter.  To assess the significance of the differences in the population medians, we also conducted Mann-Whitney U Tests \citep{mann47}.  This is a non-parametric test on the difference between two populations, with the null hypothesis being that the two populations come from the same underlying distribution.  The test statistic $U$ is calculated by ranking the pooled sample from smallest value to largest, summing the total ranks in each sample from this pooled ranking, and determining the minimum of $U_{i} = \frac{1}{2}n_{i}(n_{i}+1) - R_{i}$ between both samples, where $n_{i}$ is the size of sample $i$ and $R_{i}$ is sample $i$'s summed rank.  We provide the test statistics and associated $p-$values for these tests in the rightmost two columns of Table~\ref{tab:comp}.

To summarize Table~\ref{tab:comp}, outliers show many statistically significant differences from galaxies on the relation: they have lower $n$, lower $\Sigma_{0,\,*}$, smaller $D_{10}$, smaller $D_{\rm cluster}$, and larger $\log($PI$)$.  By contrast, our tests imply that both populations have similar colours and $R_{\rm eff}$.  The latter test is sensitive to the outlier sample, as is the difference in $n$.  The differences in medians of both properties imply the outlier dwarfs are slightly larger and less centrally concentrated than the dwarfs on the relation.

The colours of these outliers span a very limited range, with the majority of outlier colours falling between $0.7 < g^{\prime}-i^{\prime} < 1.0$, which implies stellar populations with ages $>5$~Gyr \citep[assuming reasonably low metallicities for dwarfs;][]{venhola19}.  This, in turn, suggests that recent star formation could not be the cause behind these galaxies' outlier status, as their stellar populations were most likely already in place prior to or very early into their infall in the cluster environment (though this does not rule out early star formation; we discuss this further in Sec.~\ref{sec:tides}).  Strong differences occur among environmental and structural parameters: outliers have lower central mass density (see also Fig.~\ref{fig:main} and Sec.~\ref{sec:scatter}), have smaller $n$, are nearer to their neighbours, closer to the cluster centre, and show stronger (by around a factor of 10) tidal perturbation index (though this may be attributable to their tentatively larger $R_{\rm eff}$, as previously discussed).

We have chosen outliers as those galaxies with significantly smaller $R_{3.63}$ than those on the relation.  For similar $n$, at a given $\mathcal{M}_{*}$, an increase in $R_{\rm eff}$ would decrease the surface brightness in the inner regions, moving $R_{3.63}$ inward.  However, the change in $R_{\rm eff}$ between populations is small ($\Delta R_{\rm eff} \sim 80$~pc).  We can compare how this change affects the predicted value of $R_{3.63}$ between the outlier and IRSMR populations.  Using the median values of $R_{\rm eff}$ and $n$ for the two populations for a galaxy with a stellar mass of $\log(\mathcal{M}_{*}/\mathcal{M}_{\odot}) = 7.1$ (the median mass of the outlier dwarf population), $R_{3.63}$ should decrease by $\sim25$\%, whereas the median distance from the predicted relation for the outlier population corresponds to roughly a factor of two change.  For a stellar mass of $\log(\mathcal{M}_{*}/\mathcal{M}_{\odot}) = 7.0$, the median mass of the IRSMR dwarf population, the change in $R_{3.63}$ is $\sim 60$\%.  Likewise, the difference in $\Sigma_{*,\,0}$ predicted by the change in S\'{e}rsic profile is only $\Delta\Sigma_{*,\,0} \sim 0.2 \mathcal{M}_{\odot}$~pc$^{-2}$.  Even with an additional change of $\sim0.05 \mathcal{M}_{\odot}$~pc$^{-2}$ implied by the differences in mass between the two populations (derived from the fits given in Table~\ref{tab:fits}), this is a fraction of the measured $\Delta\Sigma_{*,\,0} \sim 0.59\mathcal{M}_{\odot}$~pc$^{-2}$.

The outliers from the IRSMR thus are significantly more diffuse than those on the relation, beyond what is predicted by their differing S\'{e}rsic profiles.  We explore this structural difference in detail in following section.

\section{Discussion}\label{sec:conclusions}

\subsection{The radius-mass relation}\label{sec:irsm}

The IRSMR for dwarf galaxies has remarkably little scatter for $R_{3.63}$ (RMS$\sim 0.170$~dex; Sec.~\ref{sec:scaling}, Fig.~\ref{fig:main}).  For more massive galaxies, the low scatter in isophotal radius--stellar mass relations has long been known \citep[e.g.,][]{schombert86, binggeli91}, but recently, \citet{sanchezalmeida20} showed that it results naturally from the anti-correlation between $R_{\rm eff}$ and $n$.  For a given total magnitude (stellar mass), two surface brightness profiles which decrease monotonically eventually cross, so a radius defined close to that crossing point will show a tight relationship with magnitude (stellar mass).  A radius that minimizes the scatter in this relation can thus be found for a given population of galaxies.  For dwarfs, because they span a narrow range in both $R_{\rm eff}$ and $n$, the scatter is reduced even more than for massive galaxies, making this relation potentially extremely powerful for estimating dwarf galaxy stellar masses from single-band photometry \citep{sanchezalmeida20}, if one uses an IRSMR defined at a large enough radius to avoid the influence of outliers (Fig.~\ref{fig:alternative}).

The curvature in the relation for dwarf galaxies is distinct from that for massive ETGs, suggesting a structural dichotomy between the two mass ranges.  Massive LTGs tend to follow both the ETG relation and a rough extrapolation of the dwarf relation toward higher masses.  From the left panels of Fig.~\ref{fig:main}, many massive LTGs cluster around $n\approx1$ and $\log(\Sigma_{*,\,0})\approx3.5$, which are also rough extrapolations of both dwarf $\mathcal{M}_{*}$--$n$ and $\mathcal{M}_{*}$--$\Sigma_{*,\,0}$ relations, suggesting that the photometric profiles of Fornax Cluster dwarfs and massive LTGs share some similarities and contrast with ETGs.

\citet{derijcke09} found a similar division between dwarfs and massive galaxies, which they placed at absolute $V$-band magnitude $M_{V} = -14$.  For a colour of $B-V=0.8$, using the mass-to-light ratio conversion table from \citet{bell03} and a solar absolute magnitude of $M_{V,\odot} = 4.81$ \citep{mann15}, this corresponds to a stellar mass of $\log(\mathcal{M}_{*}/\mathcal{M}_{\odot}) \approx 8$, quite near to our boundary of $\log(\mathcal{M}_{*}/\mathcal{M}_{\odot}) = 9$.  Possibly due to a smaller sample size, they found no clear trend between $\mathcal{M}_{*}$ and $n$ for their dwarf sample, noting merely that dwarfs show values between $0.5 < n < 1.0$.  However, they did find a double-power-law correlation between $\mathcal{M}_{*}$ and central surface brightness for dwarfs, which we replicate here.

The nature of this possible structural duality has been the subject of great debate \citep[e.g.,][and many others]{kormendy85, caon93, graham03, ferrarese06, janz08, sharina08, kormendy09, misgeld09, kormendy12, calderon15}, yet, with our large, homogeneously analysed sample, we find evidence for a stark division between dwarfs and massive galaxies.  To compare with the results from \citet{derijcke09}, we used the photometric conversions from \citet{jester05} to re-derive our log-linear relation in the $V$-band, yielding a best-fit relation for the dwarf end of $\mu_{0,V} = (31.11\pm 0.44) + (0.56\pm 0.06)M_{V}$, where the uncertainties associated with each parameter include only the standard error of the fit, not errors in transforming from stellar mass to $V-$band magnitude.  This is extremely close to the relation found by \citet{derijcke09}, $\mu_{0,V} = (31.86\pm 0.43) + (0.66\pm 0.04)M_{V}$, as well as to the relation found by \citet{sharina08} for nearby dwarf irregular galaxies (with a slope of $\sim 0.5$).  The data used by \citet{derijcke09} and \citet{sharina08} included primarily data taken with the Hubble Space Telescope (with some ground-based archival imaging as well), suggesting that resolution is not a strong factor behind the appearance of this relation.  \citet{derijcke09} also derived photometric parameters by fitting S\'{e}rsic functions to the radial surface brightness profiles of their sample galaxies, a different methodology than what we employed for our sample.  The shape of this relation for dwarf galaxies thus seems robust to image resolution and fitting methodology, and possibly to environment and morphology as well.  A similar photometric analysis of a similarly large dwarf and massive galaxy sample outside of the Fornax Cluster should therefore serve to fully validate this dwarf--massive galaxy dichotomy.

In some sense, a dwarf--massive galaxy separation is expected in concordance cosmology, in which galaxy mass is built hierarchically.  In such a scenario, dwarf galaxies form first, through gas accretion into existing dark matter halos, then subsequently merge into more massive systems \citep[e.g.,][]{white78}.  Unlike stars or dark matter, gas is able to dissipate energy through brehmsstrahlung cooling \citep{hoyle53}, which is released along the rotation axis and hence flattens the system into a rotating disk; cluster dwarfs show evidence of disklike structure in their shape distributions, which imply most are oblate spheroids \citep[e.g.,][]{lisker07, sanchezjanssen16, venhola19}.  This, in turn, hearkens to the morphological sequence proposed by \citet{kormendy12}, in which dwarf spheroidal galaxies lie parallel to gas-rich dwarf irregular galaxies.  Mergers, by contrast, can create substructures within or around existing systems \citep{malin83, bullock05, cooper10, atkinson13, bilek20} and can promote the formation of instabilities \citep[e.g.,][]{toomre72, barnes92, berentzen04, athanassoula10, lokas16} which may proceed to rearrange the system's angular momentum \citep[e.g.,][]{lyndenbell72, sellwood02, athanassoula05, minchev12, donohoekeyes19}.  Indeed, the bar fraction in nearby galaxies drops precipitously below $\log(\mathcal{M}_{*}/\mathcal{M}_{\odot}) \lesssim 9.5$ \citep{diazgarcia16}, very close to where we see the transition between dwarfs and massive galaxies in these scaling relations.  \citet{su21} also found that the fraction of galaxies requiring more than one component in their decompositions drops below this mass range.  While most cosmological simulations currently have too low resolution to accurately produce dwarf galaxies at the level of detail needed to examine such scenarios, in future work, we will search for this dwarf--massive galaxy dichotomy in the high-resolution cosmological zoom simulation New Horizon \citep{dubois21}.  If found, this will help provide a theoretical foundation for our observations.

We also find clear correlations between $\mathcal{M}_{*}$ and both $n$ and $\Sigma_{*,\,0}$, such that lower mass dwarfs have lower $n$ and lower $\Sigma_{*,\,0}$.  As with many other properties of dwarfs, such as their susceptibility to tidal perturbation and RPS, this correlation may be related to the depths of their potential wells \citep[which does decline with stellar mass, even if the dark matter fraction is anti-correlated with stellar mass for dwarf galaxies, e.g., Fig.~4 of][]{moster10}.  Feedback is more effective at removing mass from or rearranging mass in low-mass dwarfs than from high-mass dwarfs, potentially leading to more diffuse structures \citep[e.g.,][]{governato10, teyssier13, dicintio17}.  Dwarfs in high-density environments in particular may be susceptible to this, as many of them show signs of rapid bursts of SF followed by quenching \citep[e.g.,][]{weisz11}, leaving no chance to build more central mass through subsequent gas accretion \citep[which is supported by simulations; e.g.,][]{joshi21}.

The lowest-mass dwarfs in our sample ($\log(\mathcal{M}_{*}/\mathcal{M}_{\odot}) < 7$) systematically lie above the IRSMR predicted from the $\mathcal{M}_{*}$--$n$ and $\mathcal{M}_{*}$--$\Sigma_{*,\,0}$ relations.  As demonstrated in Sec.~\ref{sec:scaling}, the sharp downturn in the predicted relation arises from the use of a specific isomass radius, which will sometimes be defined at a surface mass density lower than the central surface mass density predicted by the log-linear fit.  That the lowest-mass dwarfs do not show as strong a downturn as predicted is somewhat curious.  Resolution effects should act to oppose this trend: the $g^{\prime}$ and $r^{\prime}$ PSF FWHM in the FDS fields is always near $\sim$1\arcsec \ \citep[Table~A1]{venhola18}, which at the cluster's distance is $\sim 100$~pc; $R_{3.63}$ lies near this for the lowest-mass galaxies in the sample, thus may often by underestimated (meaning their true values would be even farther from the predicted curve).  The shape of the downturn, however, is also quite sensitive to the slopes in the $\mathcal{M}_{*}$--$n$ and $\mathcal{M}_{*}$--$\Sigma_{*,\,0}$ relations (Fig.~\ref{fig:theor}), so more precise estimates of light profile shapes could place more galaxies on the relation at the low-mass end.  We also assume these two relations are log-linear at the low-mass end, whereas an equally viable explanation, within the relations' scatter, is that both show some curvature and flatten at very low masses, though, as we demonstrated in Sec.~\ref{sec:outliers}, these choices have little impact on our choice of outlier population.

\subsection{The physical structure of the outliers from the IRSMR}\label{sec:alternative}

\begin{figure*}
    \centering
    \includegraphics[scale=1.0]{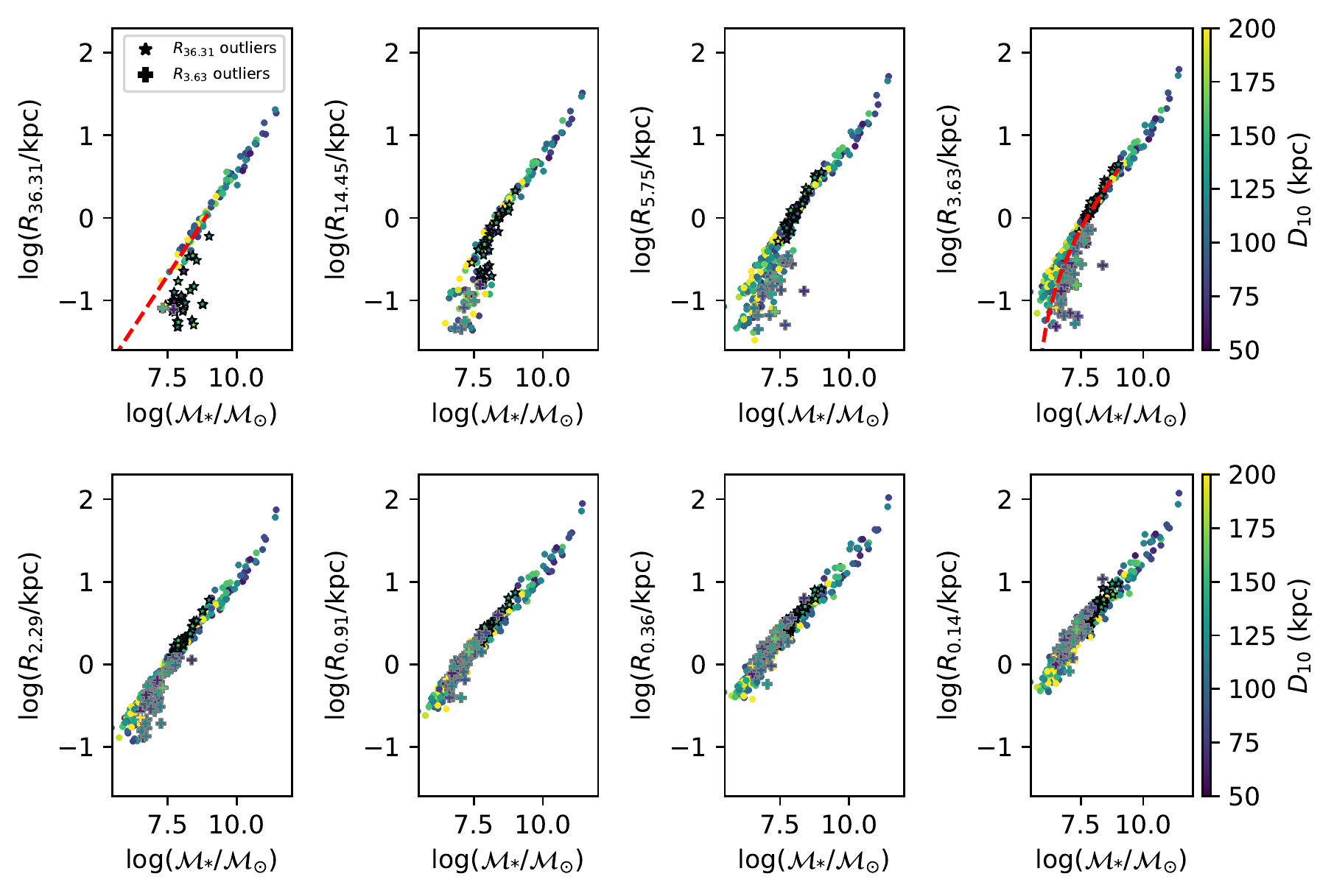}
    \caption{A range of IRSMRs for the FDS galaxies, ranging from $R_{36.31}$ (i.e., the radius corresponding to $\Sigma_{*} = 36.31 \mathcal{M}_{\odot}$~pc$^{-2}$) to $R_{0.14}$.  In each panel, we show outliers from the $R_{36.31}$ and $R_{3.63}$ relations as stars (outlined in black) and pluses (outlined in gray), respectively.  The red dashed lines in the first and fourth panels in the top row show the predicted IRSMRs used to define these outliers (dwarf galaxies only).  We colour-code the points by 10$^{\rm th}$-nearest neighbour distance.
    \label{fig:alternative}}
\end{figure*}

\begin{figure*}
    \centering
    \includegraphics[scale=1.0]{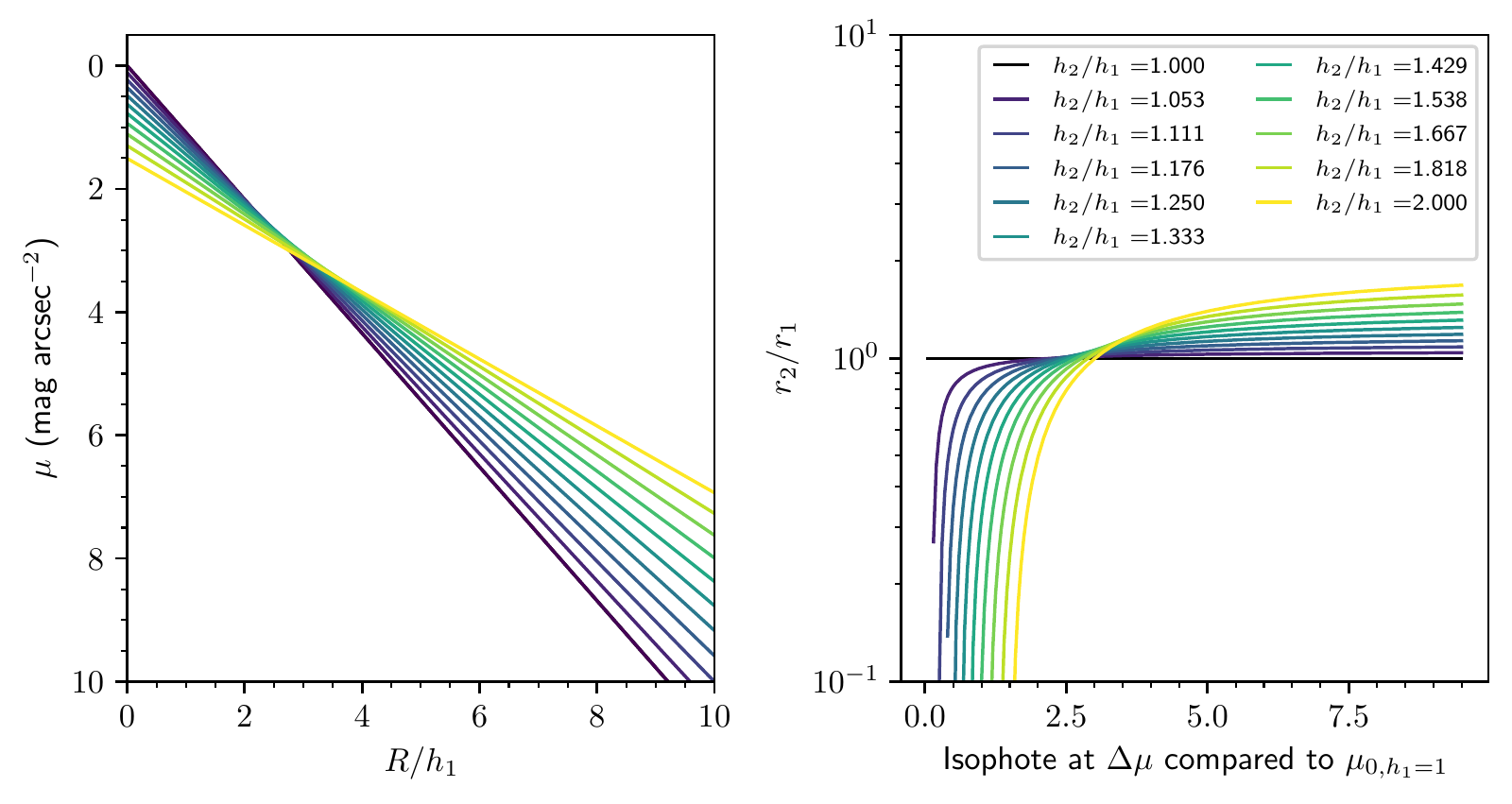}
    \caption{Demonstrating how a change in scale length affects isophotal radii.  The \emph{left} panel shows a series of exponential surface brightness profiles (arbitrary surface brightness units) with the same integrated flux, with radii normalized to the fiducial (black) profile's scale length $h_{1}$.  The curves in the \emph{right} panel show the ratios of isophotal radii defined from the enhanced scale length curves ($r_{2}$) to those of the fiducial profile ($r_{1}$) against the isophotal surface brightnesses which define these radii, relative to the central surface brightness of the fiducial profile.  We provide the scale length ratios of each curve to the fiducial in the legend.
    \label{fig:exponential}}
\end{figure*}

Here we assess what the true physical origin of the IRSMR outliers are by examining the behaviour of the IRSMR itself, and how our choice of fiducial IRSMR influences what we define as outliers.  We demonstrate the behavior of the IRSMR using different choices of isomass radius in Fig.~\ref{fig:alternative}, showing the FDS galaxies only.  In each panel, we colour-code the points by 10$^{\rm th}$-nearest neighbour distance $D_{10}$.  We show outliers from two IRSMRs (defined using the same lower mass limit and distance from the respective relation as before) using different symbols: stars (outlined in black) show outliers from the $R_{36.31}$--$\mathcal{M}_{*}$ relation (i.e., the isomass radius relation corresponding to a surface mass density of $36.31 \mathcal{M}_{\odot}$~pc$^{-2}$, equivalent to an isophotal radius of $R_{23.0,\,3.6}$; we define all other radii in a similar fashion), while we show outliers from our standard $R_{3.63}$--$\mathcal{M}_{*}$ relation as pluses (outlined in gray).  We show these two relations as dashed red lines in their corresponding panels.

Due to the correlation between $\mathcal{M}_{*}$ and $\Sigma_{0,\,*}$, low-mass galaxies drop off systematically as higher-mass isodensity radii are used.  When they are preserved, however, outliers from low-isodensity radius relations remain as outliers from high-isodensity radius relations.  Similar to our fiducial relation, we see hints that outliers from high-isodensity radius relations tend to lie in denser local environments than galaxies on their respective IRSMRs.

Additionally, outliers from relations defined using high-isodensity radii (e.g., $R_{36.31}$) tend to migrate onto the IRSMRs defined using lower-isodensity radii.  Each relation thus seems capable of probing structurally unusual galaxies within only a limited mass range.  Relations using the lowest-isodensity radii are seemingly incapable of identifying structural outliers within the range of galaxy masses probed by our catalogue: from $R_{0.91}$ (equivalent to an isophotal radius of $R_{27.0,\,3.6}$) and lower, the relation shows almost no outliers at any mass.  $R_{0.91}$ is roughly the isomass radius used by \citet{trujillo20} and \citet{chamba20}; we show here that it maintains its low scatter to much lower masses than what those two studies' catalogues were able to probe given their use of external, less low-surface-brightness-sensitive catalogues to define their sample.  Its tightness appears to arise because outliers from higher-isodensity radius relations have migrated nearly to the centre of this one; these same outliers continue migrating upward with lower-isodensity radius relations, hugging the top of the $R_{0.14}$ relation and thus contributing to their already inrinsically larger scatter \citep[see:][]{sanchezalmeida20}.  This behavior implies that a radius exists which might minimize the scatter for all stellar masses, possibly driven the physical mechanisms which limit galaxy growth \citep[e.g., galaxy truncations;][]{chamba22}.

The behavior of these outliers on these various relations can be explained by their structural differences.  We demonstrate this in Fig.~\ref{fig:exponential}.  In the left panel, we show a range of artificial exponential surface brightness profiles (with arbitrary flux units), each normalized to the scale length of the fiducial curve ($h_{1}$), shown in black.  Every curve has the same integrated flux; only the scale lengths change.  In the right panel, we show how this change in scale length affects a range of isophotal radii.  Each curve shows, as a function of isophotal surface brightness (relative to the fiducial curve's central surface brightness), the ratios of isophotal radii measured from curves with increased scale lengths ($r_{2}$) to the same isophotal radii on the fiducial curve ($r_{1}$).  We show the ordinate axis in a logarithmic scale to match the scaling we have used throughout the paper to define the IRSMR; in this scaling, a change in scale length alters isophotal radii in the profile's inner regions, inside of where the curves in the left panel cross, more severely than those in the outer regions.  The change in the inner region is also in the opposite direction to that in the outer region: for example, if a galaxy with a central surface brightness of $\mu_{0,\,3.6} = 24$ mag arcsec$^{-2}$ increases its scale length by a factor of two, while preserving its total magnitude, the initial isophote $\mu = 25.5$ mag arcsec$^{-2}$ decreases in surface brightness to 26.25 (a factor of two decrease in linear units), while the isophote $\mu = 28.0$ mag arcsec$^{-2}$ increases to 27.5 (a factor of $\sim1.6$ increase in linear units).  Likewise, the initial isophotal radius $R_{25.5}$ decreases by a factor of 1.67, while $R_{28}$ increases by a factor of 1.14.  As most dwarf galaxies have nearly exponential profiles and low central surface brightnesses (Fig.~\ref{fig:main}), this explains the behavior seen in Fig.~\ref{fig:alternative}, as well as the outliers' enhanced $R_{\rm eff}$ (for an exponential disk, $R_{\rm eff}=1.67h$ for scale length $h$).  Among the IRSMR outliers, stellar mass has seemingly migrated from their centres to their outskirts, without significant loss to the cluster potential.

\subsection{The tidal nature of the outliers from the IRSMR}\label{sec:tides}

We demonstrated in Sec.~\ref{sec:outliers} that the outliers from the IRSMR are under-dense, have larger half-light radii, and lie in relatively dense local environments, generally closer to the Fornax cluster centre.  All outliers also have smaller $R_{3.63}$ compared to those on the relation, which we have demonstrated in Sec.~\ref{sec:alternative} likely arose through a uniform expansion of the mass density profile.  Their larger $R_{\rm eff}$ also favours this explanation \citep[e.g.,][]{moore98, janz16, venhola19}.

Moving mass within a gravitationally bound system requires an injection of energy.  In galaxies, this can come from two primary sources: feedback or external perturbations.  Feedback, in the form of strong gas outflows, is often invoked as a mechanism for producing cored dark matter halos in galaxies \citep[e.g.,][]{oh11, onorbe15, dicintio17}, which also shapes the stellar structure.  This process depends on the feedback's timing in relation to the halo assembly; if it occurs before the halo has fully assembled, truncating or dying in intensity afterward, the resulting halo has a cuspier inner profile and hence smaller stellar $R_{\rm eff}$ brought about by the unimpeded subsequent dark matter assembly.  Strong outflows occurring after most of the halo has assembled produces cored profiles and extended stellar $R_{\rm eff}$ \citep[e.g.,][]{onorbe15, dicintio17}.  Fig.~3 from \citet{dicintio17} is particularly striking, as it shows an example of a galaxy with such a post-assembly starburst-fueled star-formation history whose old stellar population shows an evolution mimicking the outward migration of mass we proposed to explain the IRSMR outliers' structures in Sec.~\ref{sec:alternative}.  Strong star formation resulting in gas outflows thus appears capable of producing galaxies like the IRSMR outliers in Fornax.

However, the outliers also preferentially lie in denser local environments.  Indeed, in Fornax, dwarfs with unusually large $R_{\rm eff}$ for their stellar mass are always preferentially found nearer the cluster center \citep{venhola22}.  If, as proposed in the above scenario, star formation-induced feedback were the primary cause of these expansions, this would imply that dwarfs in dense local environments retained their gas longer than the typical galaxy found in relatively under-dense regions, e.g., nearer to the cluster edge or in the sub-group Fornax A.  This is counter-intuitive, given the stronger ram pressure and tidal forces found in this environment's densest regions, near to the cluster centre.

While the dwarfs' early star formation histories (SFHs), prior to cluster infall, likely do impact their structures to some extent, we would not expect these to correlate with the dwarfs' ultimate positions relative to other galaxies, nor their ultimate positions within the cluster itself \citep[assembly times for dwarfs with this stellar mass are between $2$ and $4$~Gyr;][]{dicintio17}.  Tidal forces, however, do correlate with local environment, therefore it is worth discussing how such forces affect the structures of low-mass galaxies.

Significant stellar mass loss is a rare event even in dense environments \citep[e.g., Fig.~4 of][]{jackson21}, requiring multiple close passages with massive companions \citep[see also][]{mastropietro05, penarrubia08}. In Fornax, the signature of mass loss is likely most clearly visible in the relative lack of low-surface-brightness galaxies in the cluster core \citep[e.g., Fig.~13 of][]{venhola17}.  The outlier galaxies in our sample, however, do not reside in the cluster core, but just outside it, suggesting they suffered far less mass loss in their evolution.  In the absence of significant mass loss, tidal perturbation is known to increase $R_{\rm eff}$, particularly in galaxies with low central densities and therefore shallow potential wells  \citep[such as low-surface-brightness galaxies;][and references therein]{martin19, jackson21}.

The scenario proposed by \citet{martin19} and \citet{jackson21} for the formation of low-surface-brightness galaxies seems particularly pertinent here: such galaxies, in the simulations they investigate, emerge from relatively dense environments and so have SFHs which peak very early, heating their dark matter haloes (and, therefore, their stellar distributions) via the mechanism discussed by, for example, \citet{dicintio17}.  At late epochs, however, their positions within dense environments, in combination with their feedback-induced shallower, cored potentials, causes them to be stripped of gas and tidally perturbed into an even more diffuse state, which cannot be reversed through gas infall \citep[as in the scenario proposed by, e.g.,][]{moore98} due to ram pressure stripping.  Fornax dwarfs in dense regions of the cluster can also be found close to the IRSMR (Fig.~\ref{fig:colored}) despite showing central mass densities similar to dwarf populations found in the cluster outskirts; it may thus be that the IRSMR outliers are primarily those dwarfs initially made less dense by vigorous early SFHs (thus more prone to additional inflation via tidal perturbation), while dwarfs in similar cluster regions with higher central mass densities had more prolonged SFHs resulting in denser, more resilient structures.

Given this, we find that tidal perturbations---possibly in combination with a particularly vigorous early SFH, though we cannot constrain this with our available data---are the most likely culprit behind the creation of the IRSMR outlier dwarfs, injecting orbital energy into the stars and dark matter through gravitational encounters \citep[however, for an intriguing counter-argument, see][]{asencio22}.  Strong central star formation likely contributed to some degree, given the outliers' more abundant nuclei, but this, too, may have been tidally triggered as the dwarfs orbited through the dense cluster potential.  If, however, vigorous early star formation were primarily at fault, this would imply a deep connection between these dwarf galaxies' pre-cluster-infall SFHs and the eventual density of their local environments, suggesting that the impacts of the cosmic web on galaxy evolution begin very early.

\subsection{Ultra-diffuse galaxies}\label{sec:udgs}

\begin{figure}
    \centering
    \includegraphics[scale=1.0]{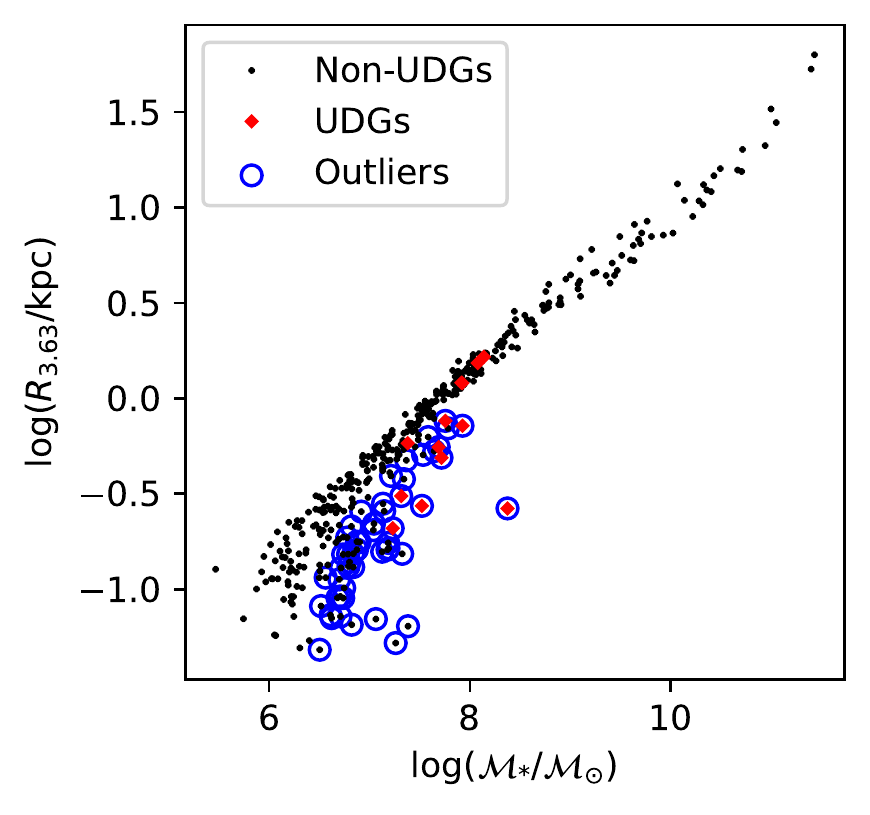}
    \caption{The IRSM for FDS galaxies, with UDGs marked by red diamonds.  IRSMR outliers are circled in blue.
    \label{fig:udgs}}
\end{figure}

Having established that the outliers from the IRSMR are likely tidally perturbed galaxies, we show the positions of the ultra-diffuse galaxies (UDGs) identified by \citet{venhola17} on the IRSMR in Fig.~\ref{fig:udgs}.  While a large fraction of these UDGs are outliers from the relation, five are not.  Of these five, two are outliers from the $R_{14.45}$ IRSMR, and one may be an outlier from the $R_{36.31}$ IRSMR, although it falls within the region of this relation with high scatter.  This suggests that UDGs in Fornax are, in general, structural outliers in the same manner as the dwarfs we have discussed throughout this paper, accounting for their diffuse appearances.  However, because the currently accepted definition of UDGs applies thresholds in $R_{\rm eff}$ and central surface brightness, and because stellar mass is correlated with both of these parameters (even among our outlier population), most of the IRSMR outliers we identified with masses below $\log(\mathcal{M}_{*}/\mathcal{M}_{\odot}) \lesssim 7.5$ are not considered UDGs.  Additionally, four outliers from the IRSMR above this mass threshold, with small clustercentric distance ($D_{\rm cluster} \sim 0.5$--$1$~deg), are not classified as UDGs.  The term UDG thus appears to mix populations of tidally perturbed dwarfs with normal dwarfs in Fornax, while excluding the bulk of the low-mass population of tidally perturbed dwarfs.

Should the term UDG continue to be employed, our analysis suggests that its definition might benefit from some revision, possibly using the IRSMR as a baseline (although we have shown that this, too, has its limitations).  For example, for galaxies with stellar masses $10^{7} < \mathcal{M}_{*}/\mathcal{M}_{\odot} < 10^{9}$, a UDG might be defined as falling more than $N\times$ the IRSMR RMS below the $R_{36.31}$, $R_{14.45}$, $R_{5.75}$, or $R_{3.63}$ IRSMRs (or using any other similar isomass radii) for some multiplicative factor $N$.  Lower-mass UDGs might be classified using isomass radii defined at fainter isodensities, assuming outliers from these relations do exist (which remains unclear given that none appear in our sample).

Some caution is merited, however, in such labeling.  As shown in Fig.~\ref{fig:main}, these outliers are primarily dwarfs occupying the low-$\Sigma_{*,\,0}$ tail for their stellar masses, which is partly why they appear as outliers in the IRSMR (Sec.~\ref{sec:scatter}).  If tidal perturbation is at fault for these low central densities, we would expect that more isolated group or field dwarf populations would show a tighter $\mathcal{M}_{*}$--$\Sigma_{*,\,0}$ relation, possibly with a shallower slope arising from systematically higher $\Sigma_{*\,0}$ values per stellar mass, and therefore would show fewer IRSMR outliers as well.  In such a scenario, one might reasonably conclude that the cluster environment broadens the $\mathcal{M}_{*}$--$\Sigma_{*,\,0}$ relation compared to its unperturbed counterpart, creating a unique class of tidally perturbed dwarfs which are rare outside of clusters.  Classifying IRSMR outliers as UDGs could be merited if this proves true.  However, if the abundance of IRSMR outliers is similar between cluster and field environments, then we must conclude that these dwarfs are merely drawn from the high-$\sigma$, low central surface brightness tail of the broader dwarf galaxy population, and would therefore merit no special moniker.

In summary, the fairly uniform, red colours, the diffuse structures, and the strong tidal influence present from the Fornax Cluster potential all point toward tidal perturbation being the most likely primary influence moving the outlier dwarfs from the IRSMR.  We have proposed that these dwarfs are most likely to be the survivors of immense tidal influence from their cluster environment, making them potentially quite valuable targets for follow-up observations and further study of the resilience and dynamics of low-mass galaxies.

\section{Summary}\label{sec:summary}

Using imaging data and derived data products from the Fornax Deep Survey and \emph{Spitzer} Survey of Stellar Structure in Galaxies, we have presented three dwarf galaxy scaling relations---stellar-mass--S\'{e}rsic index ($\mathcal{M}_{*}$--$n$), stellar-mass--central mass surface density ($\mathcal{M}_{*}$--$\Sigma_{*,\,0}$), and stellar-mass--isomass radius ($\mathcal{M}_{*}$--$R_{3.63}$)---contrasted against those of more massive galaxies.  We find that dwarf and massive galaxies occupy distinct such relations, separated at roughly $\log(\mathcal{M}_{*}/\mathcal{M}_{\odot}) \sim 9$, a distinction made quite clear by our combined catalogue's large sample size and faint limiting surface brightness.  Likewise, dwarfs appear to show log-linear $\mathcal{M}_{*}$--$n$ and $\mathcal{M}_{*}$--$\Sigma_{*,\,0}$ relations, with the former such relation only hinted at in previous surveys.  We derive the shape of the $\mathcal{M}_{*}$--$R_{3.63}$ relation using log-linear fits to these other two, and find that it matches the shape shown by the data quite well, but only if a division is made between the two mass regimes.  This provides further evidence, demonstrated in some previous studies, that dwarf and massive galaxies are distinct in terms of the way in which their light profile shapes correlate with their stellar masses.

Owing to the morphological regularity of dwarfs, the scatter in the IRSMR is extremely small for such galaxies, which implies that dwarf galaxies are structurally self-similar.  We investigated the nature of the outliers from this relation, comparing their physical parameters to those of dwarfs on the relation.  While the outliers (all of which have low relative isomass radii) have very similar integrated colours, suggesting stellar populations with ages $>5$~Gyr, we find that outliers have lower central mass concentrations and larger half-light radii than galaxies on the relation.  They are also lie in significantly denser local environments, primarily closer to the cluster centre, which is difficult to explain if their early, pre-assembly SFHs were purely to blame for their diffuse natures.  Combined, this implies that these outliers have been tidally puffed up by the general Fornax Cluster environment.  Ultra-diffuse galaxies lie both on and off of the IRSMR, and the photometric criteria defining the term exclude the bulk of the low-mass outlier population, suggesting that the term UDG is somewhat ill-adapted for identifying structurally interesting dwarfs.  Should this term continue to be employed, it might be re-defined as galaxies which fall below a range of IRSMRs, defined with isomass radii close to the centers of one's sample galaxies.

\section*{Acknowledgements}

We thank the anonymous referee for their thorough and careful assessment of the paper.  SK and AW acknowledge support from the STFC [ST/S00615X/1]. SK acknowledges a Senior Research Fellowship from Worcester College, Oxford. JHK and JR acknowledge financial support from the State Research Agency (AEI-MCINN) of the Spanish Ministry of Science and Innovation under the grant "The structure and evolution of galaxies and their central regions" with reference PID2019-105602GB-I00/10.13039/501100011033, from the ACIISI, Consejer\'{i}a de Econom\'{i}a, Conocimiento y Empleo del Gobierno de Canarias and the European Regional Development Fund (ERDF) under grant with reference PROID2021010044, and from IAC project P/300724, financed by the Ministry of Science and Innovation, through the State Budget and by the Canary Islands Department of Economy, Knowledge and Employment, through the Regional Budget of the Autonomous Community. JR acknowledges funding from University of La Laguna through the Margarita Salas Program from the Spanish Ministry of Universities ref. UNI/551/2021-May 26, and under the EU Next Generation programme. For the purpose of open access, the author has applied a Creative Commons Attribution (CC BY) licence to any Author Accepted Manuscript version arising from this submission. 

\section*{Data Availability}

The data underlying this article are available at the Centre de Donn\'{e}es Astronomique de Strasbourg (CDS), at \url{ 10.26093/cds/vizier.36470100} and \url{https://cdsarc.cds.unistra.fr/viz-bin/cat/J/A+A/660/A69}; at the NASA/IPAC Infrared Science Archive, at \url{https://irsa.ipac.caltech.edu/data/SPITZER/S4G/overview.html}; and at the European Southern Observatory Science Archive Facility, at \url{https://archive.eso.org/cms.html}.  We will also make all of our isomass radii available in table format at MNRAS and at the CDS upon publication.  We show an example of this table in Section~\ref{sec:app_tab}.



\bibliographystyle{mnras}
\bibliography{references}



\appendix

\section{Consistency across surveys}\label{sec:app_cons}

\begin{figure}
    \centering
    \includegraphics[scale=1.0]{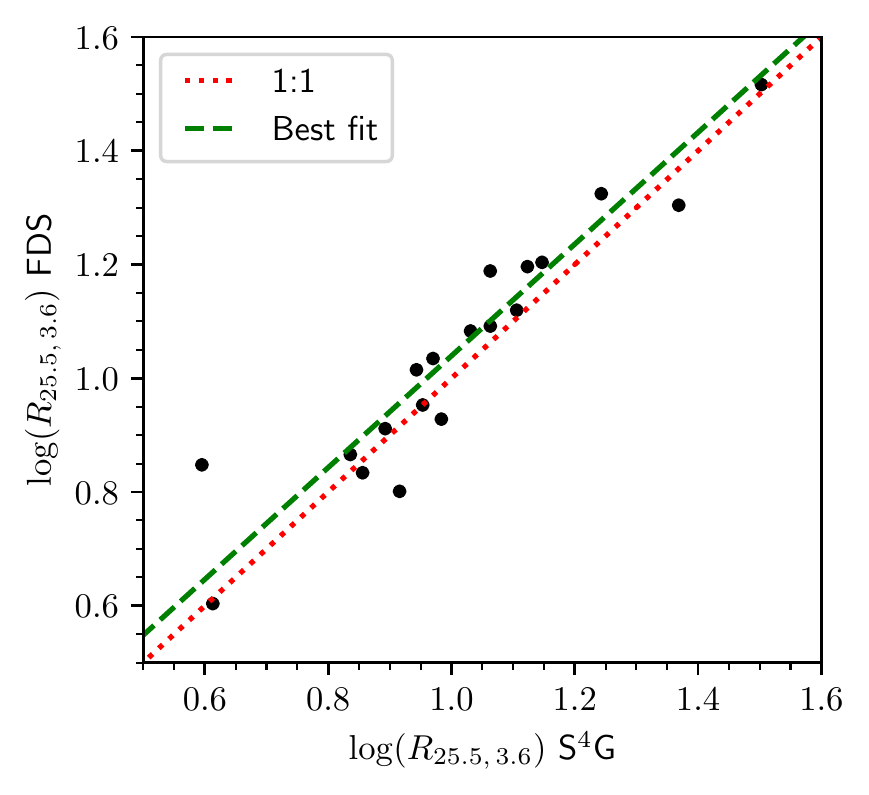}
    \caption{Isophotal radii derived for galaxies matching between the S$^{4}$G ($x-$axis) and FDS ($y-$axis).  The red dotted line gives the one-to-one relation, while the green dashed line is the best-fit linear relation.
    \label{fig:comparison}}
\end{figure}

Fig.~\ref{fig:comparison} shows the degree of agreement between isophotal radii for matching galaxies between the S$^{4}$G and FDS samples (19 in total).  The $x-$axis shows the base-10 logarithm of the $\mu_{3.6,\,AB} = 25.5$ mag arcsec$^{-2}$ isophotal radii, in kpc, derived from the S$^{4}$G images, while the y-axis shows the same but derived from the FDS images using the transformation give in Eq.~\ref{eq:mu36_r}.  We show the one-to-one relation as a red dotted line, while we show our best-fit linear relation as a green dashed line, which has the form
\begin{equation}\label{eq:agreement}
Y = (0.98\pm0.05)X + (0.06\pm0.05)
\end{equation}
where $Y$ and $X$ are the logarithms of the isophotal radii measured from the FDS and S$^{4}$G images, respectively. Within the uncertainties, this is very close to the expected one-to-one relation, showing that the radii derived from each sample are consistent.  The RMS of this fit is $0.076$~dex, around half of the RMS in the IRSMR for dwarf galaxies.

\section{Photometric and calibration errors}\label{sec:app_errs}

\begin{figure*}
    \centering
    \includegraphics[scale=1.0]{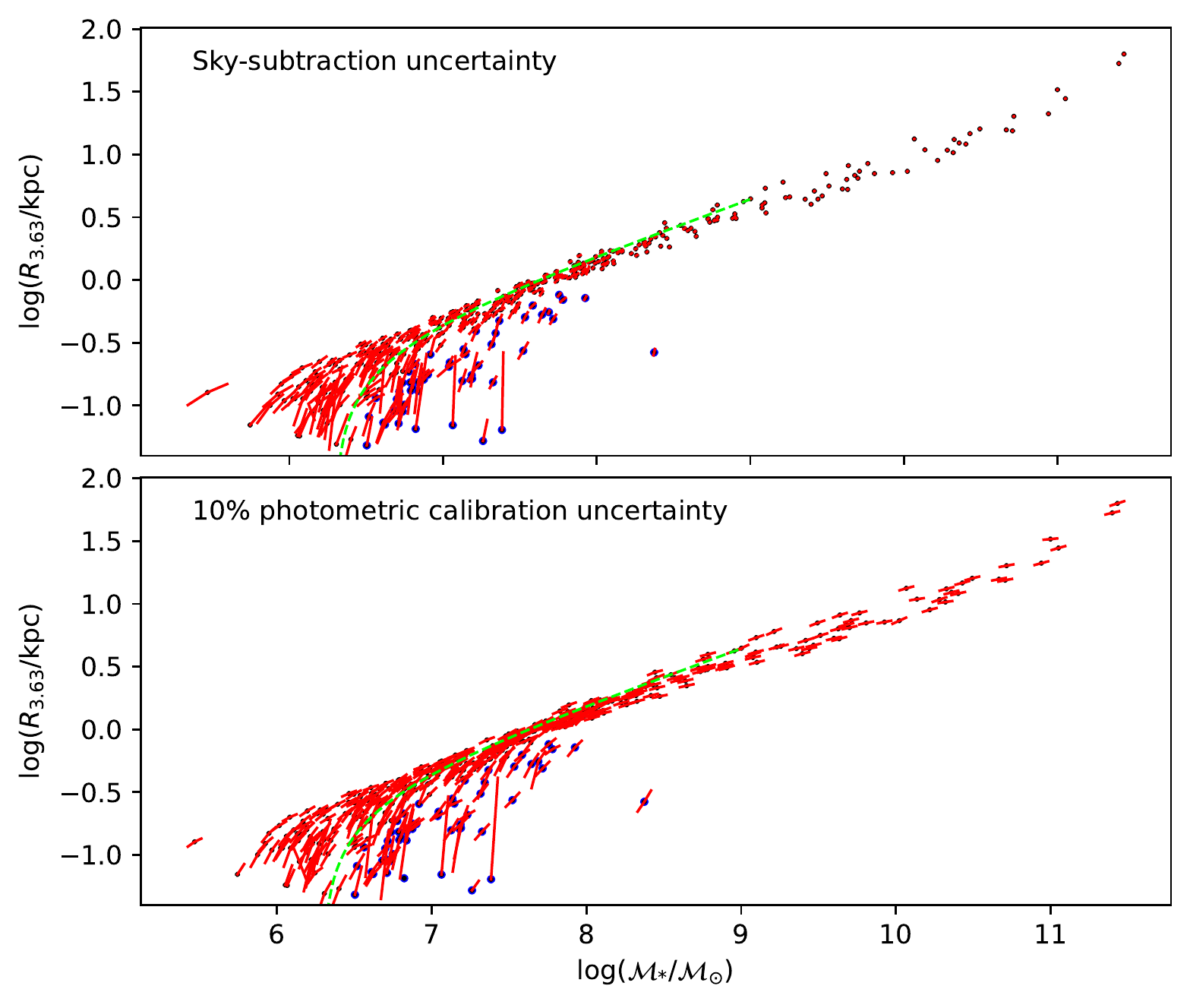}
    \caption{The IRSMR for FDS galaxies, with correlated uncertainties on stellar mass and isomass radius shown as red lines.  We show our fiducial theoretical IRSMR for dwarfs as the dashed green line, and we show our fiducial IRSMR outliers as larger blue points.  The \emph{top} panel shows uncertainties incurred from local sky-subtraction, while the \emph{bottom} panel shows the effects of the $\sim10$\% uncertainty on our conversion from FDS photometric band magnitudes to stellar mass (Eq.~\ref{eq:mstar}).  Points with no down-facing errorbars were those for which the lower estimate on $R_{3.63}$ was negative or zero.
    \label{fig:uncertainties}}
\end{figure*}

Fig.~\ref{fig:uncertainties} shows the IRSMR for FDS galaxies, with errorbars overplotted on all points as red lines.  The top panel shows uncertainties incurred due to the sky-subtraction local to the galaxies, while the bottom panel shows uncertainties incurred due to our photometric calibration, specifically the $\sim10$\% uncertainty on our conversion from $g$, $r$, and $i$ band magnitudes to stellar mass (Eq.~\ref{eq:mstar}; see Sec.~\ref{sec:methods}).  For galaxies for which the lower estimates on $R_{3.63}$ are negative, we show only upward errorbars.  Errors on stellar mass and isophotal (isomass) radius are correlated; an over-subtraction of sky, for example, reduces both the total stellar mass and the stellar mass surface density at all radii, as does a reduced estimate of stellar mass from the photometric calibration.  Errorbars are thus tilted.  This occurs in such a way that photometric uncertainties, both from measurement and from calibration, barely affect our outlier sample: when assuming upper estimates on stellar masses (thus, also on isomass radii, which are determined by the stellar masses), our sample of IRSMR outliers either does not change (from sky-subtraction uncertainty) or changes by only one (from calibration uncertainty).  Assuming instead a 30\% uncertainty (combining photometric and maximum distance-based uncertainty) reduces the outlier sample by only two.  Consequently, the effect of these uncertainties on our population comparisons (Table~\ref{tab:comp}) is negligible, and does not alter any of the conclusions presented in this paper.  Given the correlated nature of the stellar mass and isomass radius uncertainties, this is likely to be true even if our stellar mass estimates are incorrect by a considerable margin \citep[e.g., if our IRSMR outliers also happen to be outliers in mass-to-light ratio; see Fig.~12 from][]{taylor11}, though without more robust estimates of stellar mass, the full impact of this will remain uncertain.

\section{Isophotal radius table}\label{sec:app_tab}

\begin{table*}
\centering
\caption{Subset of isomass radii for FDS sample, the full table of which is available online.  We include FDS identifiers, galaxy central coordinates in decimal degrees, stellar masses and uncertainties in $\log(\mathcal{M}_{\odot})$ units, and a range of isomass radii in arcseconds. Radius column headers are labeled R followed by the mass surface density defining the radii (e.g., R36.31 is the radius corresponding to 36.31 $\mathcal{M}_{\odot}$~pc$^{-2}$).  We will also include fiducial isomass radii $R_{3.63}$.  We denote missing values by $-999$.  In total, the table contains 17 columns (nine isomass radii, down to a density of 0.14 $\mathcal{M}_{\odot}$~pc$^{-2}$) and 594 rows (excluding the header).}
\label{tab:rads}
\begin{tabular}{ccccccccccc}
\hline\hline \\
Name & RA & Dec & log(Mstar) & dlog(Mstar) & R36.31 & R14.45 & R5.75 & R3.63 & R2.29 & ... \\
\hline \\
FDS10\textunderscore0003 & 54.222 & -34.938 & 7.882 & 0.004 & 0.491 & 4.687 & 11.129 & 13.988 & 16.694 & ...  \\
FDS10\textunderscore0004 & 55.169 & -34.949 & 6.918 & 0.024 & -999 & -999 & -999 & 2.629 & 5.998 & ...  \\
FDS10\textunderscore0014 & 54.347 & -34.900 & 8.166 & 0.004 & -999 & 6.814 & 14.027 & 17.752 & 21.764 & ... \\
FDS10\textunderscore0017 & 54.593 & -34.927 & 6.307 & 0.073 & -999 & -999 & -999 & 0.508 & 1.288 & ...  \\
FDS10\textunderscore0023 & 54.478 & -34.882 & 7.734 & 0.007 & -999 & 4.781 & 9.057 & 11.532 & 14.516 & ...  \\
... & ... & ... & ... & ... & ... & ... & ... & ... & ... & ... \\
\hline
\end{tabular}
\end{table*}

Table~\ref{tab:rads} shows a subset of the parameters we will provide for download.  This includes galaxy names, coordinates, stellar masses (and associated uncertainties, both in solar units), as well as isomass radii in arcseconds corresponding to $3.6\mu$m surface brightnesses of $23$, $24$, $25$, $25.5$, $26$, $26.5$, $27$, $28$, and $29$, which correspond to $\Sigma_{*} = 36.31$, $14.45$, $5.75$, $3.63$, $2.29$, $1.45$, $0.92$, $0.36$, and $0.14$, respectively.  Additionally, we will provide the sky-subtraction uncertainties and 10\% calibration uncertainties on our fiducial isomass radii $R_{3.63}$.


\bsp	
\label{lastpage}
\end{document}